\documentclass[12pt]{article}
\usepackage{graphicx}
\usepackage{float}
\usepackage{makecell} %multiple line in a table cell
\usepackage{soul} %highlight text
\usepackage{pbox} %multiple line in a table cell
\usepackage{titling} %for adding supp
\usepackage{multirow}

\usepackage{subcaption}
\usepackage[dvipsnames]{xcolor}
\usepackage{stmaryrd}
\usepackage{todonotes}
\usepackage[skip=10pt plus1pt, indent=20pt]{parskip}
\usepackage{amsmath}
\usepackage{amssymb}
\usepackage{indentfirst} %indent after a section
\usepackage{geometry}
\usepackage[utf8]{inputenc}
\usepackage{lscape} %landscape table
\usepackage{float} %figure position fixed
\usepackage{tensor}%reference over figure
 \usepackage[percent]{overpic}  %reference over figure
\usepackage[T1]{fontenc}%reference over figure
\usepackage[export]{adjustbox}
\usepackage{tikz-cd}%reference over figure
\usepackage[outline]{contour}%reference over figure
\usepackage{lipsum,tikz}%reference over figure
\usepackage{hyperref}
\DeclareMathOperator*{\minimise}{minimise}
\usepackage[backend=bibtex,style=numeric,natbib=true,sorting=none]{biblatex} % Use the bibtex backend with the 'numeric-comp'/'authoryear' citation style (which resembles APA) (Template)
\usepackage{rotating} %sideways table
\usepackage{tabularx,ragged2e,booktabs}
\usepackage{authblk}
\usepackage{tcolorbox} % create box
\usepackage[ruled,vlined,linesnumbered,resetcount]{algorithm2e}  
\newcommand{\pv}{\textit{P. vivax}}
\newcommand{\pf}{\textit{P. falciparum}}
\newcommand{\etal}{\textit{et al. }}

\begin{document}

\title{Investigation of \textit{P. Vivax} Elimination via Mass Drug Administration}

\author[1*]{Md Nurul Anwar}
\author[1,2]{James M. McCaw}
\author[1]{Alexander E. Zarebski}
\author[1,3,4,\dag]{Roslyn I. Hickson}
\author[1,\dag]{Jennifer A. Flegg}

\affil[1]{School of Mathematics and Statistics, The University of Melbourne, Parkville, Australia}
\affil[2]{Centre for Epidemiology and Biostatistics, Melbourne School of Population and Global Health, The University of Melbourne, Parkville, Australia}

\affil[3]{Australian Institute of Tropical Health and Medicine, James Cook University, Townsville, Australia}
\affil[4]{CSIRO, Townsville, Australia}

\date{}                     %% if you don't need date to appear
\setcounter{Maxaffil}{0}
\renewcommand\Affilfont{\itshape\small}
\maketitle
* nurul.anwar@unimelb.edu.au\ 
\dag\ \small{These authors contributed equally to this work}
\begin{abstract}
\textit{Plasmodium vivax} is the most geographically widespread malaria parasite due to its ability to remain dormant (as a hypnozoite) in the human liver and subsequently reactivate. 
Given the majority of \textit{P. vivax} infections are due to hypnozoite reactivation, targeting the hypnozoite reservoir with a radical cure is crucial for achieving \textit{P. vivax} elimination. 
Stochastic effects can strongly influence dynamics when disease prevalence is low or when the population size is small. Hence, it is important to account for this when modelling malaria elimination.
We use a stochastic multiscale model of \textit{P. vivax} transmission to study the impacts of multiple rounds of mass drug administration (MDA) with a radical cure, accounting for superinfection and hypnozoite dynamics.
Our results indicate multiple rounds of MDA with a high-efficacy drug are needed to achieve a substantial probability of elimination.
This work has the potential to help guide \textit{P. vivax} elimination strategies by quantifying elimination probabilities for an MDA approach.
\end{abstract}


\section{Introduction}\label{Intro}
Among the five species of the \textit{Plasmodium} parasite, \pv~is the most geographically widespread and causes significant global morbidity and mortality \parencite{antinori2012biology,battle2019mapping}.
\textit{P. vivax} has emerged as the dominant species in Southeast Asia and was responsible for \(46\%\) of cases (5.2 million total) in 2022 \cite{WHO2023}. 
An important characteristic of \pv~is its ability to remain dormant in the human liver as a \emph{hypnozoite}.
\pv~can remain dormant for up to a year, before reactivating and potentially causing onward transmission \cite{imwong2007relapses,thriemer2021towards}. 
Relapse events (where the hypnozoites reactivate) are responsible for more than \(80\%\) of \pv~infections (in the absence of radical cure treatment) \cite{robinson2015strategies}.

Despite the clinical significance of relapse, there is still uncertainty regarding the causes of hypnozoite reactivation.
Some factors thought to be relevant include fever caused by other infections (such as \pf), and recognition by the immune system of the same \textit{Anopheles} specific protein (found in the salivary glands of adult female mosquitoes) \cite{mueller2009key,hulden2011activation,white2014modelling}. 

It can be challenging to distinguish relapse from other types of recurrent malaria, such as reinfection or recrudescence.  This can be due to incomplete elimination of a blood-stage infection, often associated with treatment failure \cite{ghosh2020mathematical}.  Antimalarial drugs refer to those that clear either blood-stage or liver-stage parasites, with the specific recommended drugs depending on the parasite species. 

Implementation of radical cure treatment is a part of standard case management in all \pv~endemic settings. 
Targeting the hypnozoite reservoir is crucial in controlling or eliminating \pv, as transmission can be re-established from the reactivation of hypnozoites \cite{white2014modelling}. 
The 8-aminoquinoline class of drugs, such as primaquine and tafenoquine, clear hypnozoites from the liver, and are referred to as \emph{radical cure} drugs \parencite{wells2010targeting,taylor2019short,poespoprodjo2022supervised}. 
The current treatment recommended by the WHO for \textit{P. vivax} malaria is a combination of two antimalarial drugs: either chloroquine or artemisinin combination therapy (ACT) to clear parasites from the blood and either primaquine or tafenoquine to clear hypnozoites from the liver. 

Mass drug administration (MDA) is a strategy used to control and eliminate malaria. MDA involves treating the entire at-risk population, or a well-defined sub-population, in a location with antimalarial drugs (depending on the malaria species), regardless of whether they have symptoms or a positive malaria diagnosis \cite{newby2015review,hsiang2013mass}.  
In a radical cure MDA intervention (for \pv), individuals are typically given a combination of two drugs in line with the WHO-recommended treatment; one drug targets the blood-stage parasites and the other targets parasites in the liver. 
These radical cure MDA approaches aim to reduce both the blood-stage parasites and the size of the hypnozoite reservoir.
However, there are costs to indiscriminant drug administration; primaquine and tafenoquine can cause life-threatening \emph{haemolysis} in individuals with G6PD deficiency.
G6PD deficiency is an enzymopathy affecting up to \(30\%\) of individuals in malaria-endemic regions \cite{recht2018use}.
Therefore, G6PD testing is recommended before administering 8-aminoquinoline.

Mathematical modelling has been widely used to understand the transmission of malaria, particularly \pf, and the likely impact of interventions. 
\pv~transmission differs from \pf~transmission, in that there are recurrent infections due to hypnozoite reactivation \cite{white2014modelling,anwar2022multiscale,anwar2023optimal}.
Mathematical models have been developed to study different aspects of \pv~dynamics\cite{anwar2023scoping}: e.g. variation in hypnozoite numbers between infectious mosquito bites, the acquisition of immunity, superinfection (multiple simultaneous infections), and the effects of treatment.
Many of the mathematical models use differential equations, as more analytical methods can be brought to bear on them, relative to stochastic or agent-based models.

When exploring disease elimination scenarios, stochastic effects can be important \parencite{henle2004role, ludwig1999meaningful}.
Furthermore, when the population size is small, or the disease is at low prevalence, a stochastic model can provide more realistic representations of the transmission dynamics than deterministic models \parencite{allen2000comparison,beran1994statistics}.
We have previously modelled hypnozoite acquisition, population dynamics and \pv~transmission in a deterministic multiscale framework \cite{anwar2022multiscale}.
We also modelled the effect of the drug on both hypnozoite acquisition and infection, accounting for superinfection \cite{anwar2023optimal}.
To the best of our knowledge, no other stochastic multiscale model has been developed that can consider \textit{P. vivax} elimination while explicitly accounting for superinfection and the effects of multiple rounds of MDA on hypnozoite dynamics and \textit{P. vivax} infection \cite{anwar2023scoping}. 
In this paper, we describe a stochastic multiscale model and demonstrate how this model can be used to compute the probability of \pv~elimination under multiple MDA rounds.

\section{Methods} \label{method}

\subsection{Transmission model}\label{stochastic_model}
We partition the human population based on individuals' \pv~status.
Let $S$, $I$, and $L$ represent the number of people in the human population who are \emph{susceptible} to infection with no hypnozoites, \emph{blood-stage infected} (with or without hypnozoites), and those who are\emph{blood-stage negative but hypnozoite positive}, respectively.

While members within the $I$ or $L$ class may differ in the number of hypnozoites they have, we do not track this in our representation of the population. 
Instead, we model the distribution of the number of hypnozoites across the individuals in each class. 
This distribution is based on a stochastic, within-host model \cite{mehra2022hypnozoite}.
We assume a constant size for the human population, i.e. $T_h=S+I+L$ is constant, where $T_h$ is the human population size.

\begin{figure}[!ht]
\centering
  \includegraphics[width=130mm]{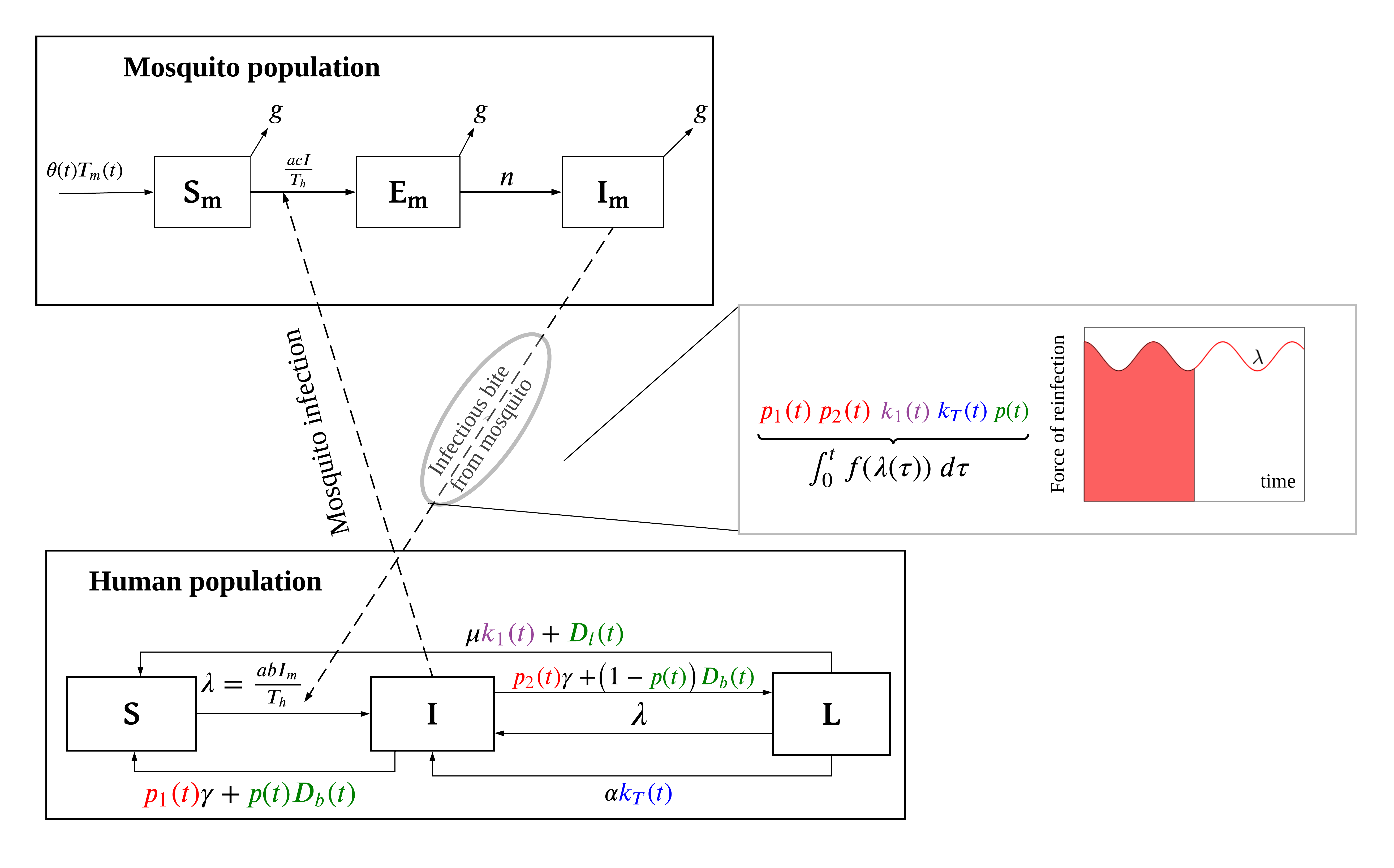}
  \caption{\textit{Schematic illustration of the multiscale model with treatment. $S$, $I$ and $L$ represent the number of the human population that are susceptible with no hypnozoites, blood-stage infected (with or without hypnozoites), and blood-stage negative but hypnozoite positive, respectively. The left (top and bottom) part of the schematic demonstrates the transmission dynamics between the human and mosquito populations. The right part of the schematic demonstrates how the within-host model (see Supplementary Material) has been embedded within the population scale model. The within-host model takes into account the history of infective bites and calculates the probability of individuals in the $I$ compartment having no hypnozoites and one blood-stage infection ($p_1(t)$), individuals in the $I$ compartment having at least one hypnozoite and one blood-stage infection ($p_2(t)$), individuals in the $L$ compartment having one hypnozoite ($k_1(t)$), the expected size of the hypnozoite reservoir ($k_T(t)$), and the probability of individuals in $I$ having no hypnozoites ($p(t)$) at any given time $t$ as a function of the force of reinfection, $\lambda(t)$. We account for superinfection through the parameters $p_1(t)$ and $p_2(t)$. The functions $D_b(t)$ and $D_l(t)$ capture the effect of treatment when implemented. Other parameters are defined in Table~\ref{tab:white}.}}
  \label{fig:multiscale}
\end{figure}

Let $S_m$, $E_m$ and $I_m$ represent the number of mosquitoes that are susceptible, exposed, and infectious, where $S_m,\ E_m,\ I_m\in \{0,1,2,\ldots,T_m\}$ and \(S_m+E_m+I_m=T_m\). 
$T_m$ is not fixed and varies seasonally due to a varying birth rate, $\theta(t)$, given by
\begin{align}
\label{eqn:mosq_seas}
    \theta(t)=g\left(1+\eta \cos\left(\frac{2\pi t}{365}\right)\right),
\end{align}
where $g$ is the baseline mosquito birth (and death) rate and $\eta\in[0\ 1)$ is the seasonal amplitude. A schematic of the stochastic multiscale model is provided in Figure \ref{fig:multiscale}.
Table~\ref{tab:Stoch} describes the possible compartment transitions and their associated rates, and Table~\ref{tab:white} describes the value and source of the parameter values used.  

\begin{landscape}
\begin{table}[]
\centering
\footnotesize
\caption{Table of transition rates and stoichiometries for the stochastic multiscale model\label{tab:Stoch}.}
\begin{tabular}[t]{m{25em}m{10cm}m{3cm}}
\toprule
\textbf{Event} & \textbf{Stoichiometries} &  \textbf{Rate} \\
\midrule
Infection of individual in $S$ compartment & $(S,\ I,\ L,\ S_m,\ E_m,\ I_m)\to (S-1,\ I+1,\ L,\ S_m,\ E_m,\ I_m)$& $abSI_m/T_h$ \\
\hline
Infection of individual in $L$ compartment & $(S,\ I,\ L,\ S_m,\ E_m,\ I_m)\to (S,\ I+1,\ L-1,\ S_m,\ E_m,\ I_m)$& $abLI_m/T_h$ \\
\hline
Relapse of individual in $L$ compartment & $(S,\ I,\ L,\ S_m,\ E_m,\ I_m)\to (S,\ I+1,\ L-1,\ S_m,\ E_m,\ I_m)$& $\alpha k_T(t)L$ \\
\hline
Death of last hypnozoite in individual in $L$ compartment & $(S,\ I,\ L,\ S_m,\ E_m,\ I_m)\to (S+1,\ I,\ L-1,\ S_m,\ E_m,\ I_m)$& $\mu k_1(t)L$\\
\hline
Natural recovery from $I$ compartment without hypnozoites & $(S,\ I,\ L,\ S_m,\ E_m,\ I_m)\to (S+1,\ I-1,\ L,\ S_m,\ E_m,\ I_m)$& $p_1(t)\gamma I$ \\
\hline
Natural recovery from $I$ compartment (with hypnozoites) & $(S,\ I,\ L,\ S_m,\ E_m,\ I_m)\to (S,\ I-1,\ L+1,\ S_m,\ E_m,\ I_m)$& $p_2(t)\gamma I$ \\
\hline
Recovery from $I$ compartment (without hypnozoites) due to radical cure  & $(S,\ I,\ L,\ S_m,\ E_m\ I_m)\to (S+1,\ I-1,\ L,\ S_m,\ E_m,\ I_m)$& $p(t)D_b(t)I$\\
\hline
Recovery from $I$ compartment (with hypnozoites) due to radical cure  & $(S,\ I,\ L,\ S_m,\ E_m\ I_m)\to (S,\ I-1,\ L+1,\ S_m,\ E_m,\ I_m)$& $(1-p(t))D_b(t)I$\\
\hline
Recovery from $L$ compartment due to radical cure & $(S,\ I,\ L,\ S_m,\ E_m\ I_m)\to (S+1,\ I,\ L-1,\ S_m,\ E_m,\ I_m)$& $D_l(t)L$\\
\hline
Birth of mosquitoes & $(S,\ I,\ L,\ S_m,\ E_m,\ I_m) \to (S,\ I,\ L,\ S_m+1,\ E_m,\ I_m)$& $\theta(t)T_m(t)$ \\
\hline
Mosquito exposure to sporozoite & $(S,\ I,\ L,\ S_m,\ E_m,\ I_m)\to (S,\ I,\ L,\ S_m-1,\ E_m+1,\ I_m)$ & $acS_mI/T_h$ \\
\hline
Mosquito becomes infectious & $(S,\ I,\ L,\ S_m,\ E_m,\ I_m)\to (S,\ I,\ L,\ S_m,\ E_m-1,\ I_m+1)$ & $nE_m$ \\
\hline
Death of susceptible mosquito & $(S,\ I,\ L,\ S_m,\ E_m,\ I_m) \to (S,\ I,\ L,\ S_m-1,\ E_m,\ I_m)$& $gS_m$ \\
\hline
Death of exposed mosquito & $(S,\ I,\ L,\ S_m,\ E_m,\ I_m) \to (S,\ I,\ L,\ S_m,\ E_m-1,\ I_m)$& $gE_m$ \\
\hline
Death of infectious mosquito & $(S,\ I,\ L,\ S_m,\ E_m,\ I_m) \to (S,\ I,\ L,\ S_m,\ E_m,\ I_m-1)$& $gI_m$ \\
\bottomrule
\end{tabular}
\end{table}
\end{landscape}

\begin{landscape}
\begin{table}[h]
\centering
\footnotesize
\caption{Definitions, values and sources for model parameters. The parameter ranges indicated in square brackets were used in the sensitivity analysis.}\label{tab:white}
\begin{tabular}[t]{clcc}
\toprule
\textbf{Symbol} & \textbf{Definition} &  \textbf{Value/s} & \textbf{Source}  \\
\midrule
$a$ & Biting rate of mosquitoes & 80  year$^{-1}$ & \parencite{garrett1964human} \\
% \hline
$b$ & Transmission probability: mosquito to human & 0.5& \parencite{smith2010quantitative} \\
% \hline
$c$ & Transmission probability: human to mosquito & 0.23& \parencite{bharti2006experimental}\\
% \hline
$\theta(t)$ & Mosquito birth rate (seasonal) & Time-varying &  Equation \eqref{eqn:mosq_seas}  \\
$g$ & Baseline mosquito birth (and death) rate & 0.1 day$^{-1}$& \parencite{gething2011modelling}   \\
% \hline
$\eta$ & Seasonal amplitude & 0.1 & Assumed \\ 
% \hline
% $\phi$ & Seasonal phase & 0 & Assumed\\ 
$T_h$ & Population size of human & 10,000& Assumed \\ 
$T_m(t)$ & Population size of mosquito& Seasonal & Modelled \\
$T_m(0)$ & Initial number of mosquito population& Calculated & $T_m(0)=m_0 T_h$ \\ 
% \hline
$m_0$ & Initial mosquito to human ratio & Varied &   \\ 
% \hline
$n$ & Rate of mosquito sporogony & 1/12 days$^{-1}$ & \parencite{gething2011modelling} \\
% \hline
$\gamma$ & Recovery rate from $I$ compartment & 1/60 day$^{-1}$& \parencite{collins2003retrospective} \\
% \hline
$\alpha$ & Hypnozoite activation rate & 1/332 day$^{-1}$ [0, 1/100]& \parencite{white2014modelling}\\
% \hline
$\mu$ & Hypnozoite death rate & 1/425 day$^{-1}$ [0, 1/100]& \parencite{white2014modelling} \\
% \hline
$\nu $ & Average number of hypnozoites per mosquito bite & 8.5 & \parencite{white2014modelling}\\
% \hline
$\lambda(t)$ & Force of reinfection & Calculated & $\lambda(t)=abI_m/T_H$\\
% \hline
% \hline
$p(t)$ &  Probability that individual in $I$ has no hypnozoites within liver & Calculated & Equation \eqref{eqn:p} \\
$p_1(t)$ & Probability that individual in $I$ has no hypnozoites and $MOI=1$& Calculated & Equation \eqref{eqn:p_1}\\
$p_2(t)$ & Probability that that individual in $I$ has hypnozoites and $MOI=1$ & Calculated & Equation \eqref{eqn:p_2}\\
% \hline
$k_1(t)$ & Probability that that individual in $L$ has 1 hypnozoite within liver & Calculated & Equation \eqref{eqn:k1}\\
% \hline
$k_T(t)$ & Average number of hypnozoites within liver for individual in $L$ & Calculated & Equation \eqref{eqn:kT}\\
% \hline
$p_{\text{blood}}$ & Probability that ongoing blood-stage infections are cleared instantaneously due to radical cure
 & 0.9 [0.5, 1]& Assumed\\
% \hline
$p_{\text{rad}}$ & Probability that hypnozoites dies instantaneously due to radical cure& 0.9 [0.5, 1]& \parencite{nekkab2021estimated} \\
% \hline
$D_b(t)$ & Clearance rate of blood-stage parasite due to radical cure& Calculated& Equation \eqref{eqn:D_b}\\
% \hline
$D_l(t)$ & Clearance rate of liver-stage parasite (hypnozoite) due to radical cure& Calculated & Equation \eqref{eqn:D_l}\\
% \hline
${N_{\text{MDA}}}$ & Total number of MDA rounds & Varied & \\
\bottomrule
\end{tabular}
\end{table}
\end{landscape}

Upon being bitten by an infected mosquito, humans in the $S$ and $L$ compartments transition to the blood-stage infected compartment ($I$).
The rate at which individuals from $S$ and $L$ transition to $I$ is $\lambda(t)=abI_m/T_h$, where $a$ is the mosquito biting rate and $b$ is the probability of transmission from a mosquito bite. 

To capture the hypnozoite dynamics and the variation of hypnozoites within individuals at the population level, we embed the within-host model of Mehra \textit{et al.} \parencite{mehra2022hypnozoite} into our model to capture the additional structure within the $I$ and $L$ compartments. We use a mean-field approximation to obtain the probability distribution of hypnozoites within individuals. 
Here, we consider the short latency case, where hypnozoites can activate immediately following establishment. 
We assume the number of hypnozoites introduced by an infectious bite follows a geometric distribution with mean $\nu$ and the dynamics of each hypnozoite are independent and identically distributed. Let $H$, $A$, $C$, and $D$ represent states of establishment, activation, clearance (removal after activation) and death (removal before activation) of a hypnozoite, respectively.
Each hypnozoite has two possible final states: death before activation $(D)$; or clearance after activation $(C)$. Furthermore, let $N_f(t)$ denote the number of hypnozoites in states  $f\in F:= \{H,A,C,D\}$ at time $t$ and $N_P(t),\ N_{PC}(t)$ denote the number of ongoing and cleared primary infections (an infection from an infectious mosquito bite), respectively, at time $t$. We then calculate the probability generating function (PGF) of $N_f,\ f\in F':=\left\{H,A,C,D,P,PC\right\}$ for the distribution of hypnozoites at time $t$ for different states (see Supplementary Material for details). 

Individuals in $I$ transition to $S$ at rate $p_1(t)\gamma$.
Here, $\gamma$ is the natural recovery rate from a blood-stage infection and $p_1(t)$ is the probability that a blood-stage infected individual has no hypnozoites and only a single blood-stage infection (i.e. they do not have a superinfection).
The derivation of \(p_1\) is given in the Supplementary Material and results in
\begin{align}
p_1(t)=&\frac{P\big(N_A(t)+N_P(t)=1|N_H(t)=0\big)P(N_H(t)=0)}{1-P(N_A(t)+N_P(t)=0)},
             \label{eqn:p_1}
\end{align}
where $N_H(t),\ N_A(t),\text{and}\ N_P(t)$ are the number of established hypnozoites in the liver, the number of relapses (that is, hypnozoites that have reactivated), and the number of primary infections, respectively.
Individuals transition from $I$ to $L$ at the rate $p_2(t)\gamma$.
Again, \(\gamma\) is the rate of natural recovery from a blood-stage infection, and now $p_2(t)$ is the probability that a blood-stage infected individual has at least one hypnozoite and only one blood-stage infection (i.e. they do not have a superinfection.)
As derived in the Supplementary Material, the equation for \(p_2\) is
\begin{align}
p_2(t)=&\frac{P\big(N_A(t)+N_P(t)=1\big)}{1-P(N_A(t)+N_P(t)=0)}-p_1(t).\label{eqn:p_2}
\end{align}

Furthermore, individuals transition from $L$ to $S$ at rate $\mu k_1(t)$ where $\mu$ is the hypnozoite death rate and $k_1(t)$ is the probability that an individual in the $L$ compartment has only a single hypnozoite:
\begin{align}
k_1(t)=&\frac{P(N_H(t)=1|N_A(t)=N_p(t)=0)}{1-P(N_H(t)=0|N_A(t)=N_P(t)=0)}.
 \label{eqn:k1}
 \end{align}

The rate individuals transition from $L$ to $I$ is $\alpha k_T(t)$, where $\alpha$ is the hypnozoite activation rate and $k_T(t)$ is the expected size of the hypnozoite reservoir within an individual in the $L$ compartment. 
The expected size of the hypnozoite reservoir is given by
\begin{align}
\label{eqn:kT}
   k_T=\sum_{i=1}^\infty ik_i &= \Big(\frac{\mathbb{E}\left[N_H(t)|N_A(t)=N_P(t)=0\right]}{1-P(N_H(t)=0|N_A(t)=N_P(t)=0)}\Big),
\end{align}
where $\mathbb{E}\left[N_H(t)|N_A(t)=N_P(t)=0\right]$ is the expected size of the hypnozoite reservoir in an uninfected (no blood-stage infection) individual. 

\subsection{Treatment}

We assume drug treatment is administered at times $t=s_1, s_2,\ldots, s_{N_{\text{MDA}}}$, where ${N_{\text{MDA}}}$ is the total number of MDA rounds.
Upon administration of the radical cure treatment, we assume the following: the blood-stage infections in an individual are instantaneously cleared with probability $p_{\text{blood}}$; and each hypnozoite in the liver dies (independently) with probability $p_{\text{rad}}$.
Therefore, whenever radical cure treatment is administered, individuals in the $I$ compartment recover with probability $p_{\text{blood}}$ and, depending on the hypnozoite reservoir, transition to compartment ($S$) or compartment $(L)$.
We define a probability $p(t)$, that a blood-stage infected individual has no hypnozoites in the liver immediately after the treatment:
\begin{align}
\label{eqn:p}
  p(t)&=P\big(N_H(t)=0|N_A(t)>0 \cup N_P(t)>0\big)\nonumber\\
  &=\frac{P(N_H(t)=0)-P(N_H(t)=N_A(t)=N_P(t)=0)}{1-P(N_A(t)=N_P(t)=0)}.
\end{align}

Assuming there is an MDA at time $t$, blood-stage infected individuals will transition to the susceptible compartment ($S$) with probability $p(t)D_b(t)$ and will transition to the $L$ compartment with probability $(1-p(t))D_b(t)$. 
Furthermore, individuals in the $L$ compartment will transition to the compartment ($S$) at an impulse $D_l(t)$ due to the radical cure.  The equations for $D_b(t)$ and $D_l(t)$ can be written as
\begin{align}
\label{eqn:D_b}
D_b(t)=&p_{\text{blood}}\sum_{j=1}^{N_{\text{MDA}}} \delta_D(t-s_j),\\
\label{eqn:D_l}
 D_l(t)=&\big\{k_1(t)p_{\text{rad}}+k_2(t)p_{\text{rad}}^2+\ldots\big\}\sum_{j=1}^{N_\text{MDA}} \delta_D(t-s_j),
\end{align}
where $\delta_D(\cdot)$ is the Dirac delta function and $s_j,\ j=1,2,\ldots, {N_{\text{MDA}}}$, are the MDA administration times. 
That is, the functions $D_b(t)$ and $D_l(t)$ take effect only at the MDA administration time. As each hypnozoite will die with the probability \(p_{\text{rad}}\) due to the effect of radical cure drug, the liver-stage clearance impulse, $D_l(t)$, depends on how many hypnozoites are present in the liver. The time-dependent probabilities $p(t),\ p_1(t),\ p_2(t),\ k_1(t)$, and $k_T(t)$ depend on the history of past infections (see Supplementary Material).

\subsection{Objective of the study} \label{obj}

Our primary objective here is to study the effect of a radical cure-based MDA intervention on the probability of \pv~elimination. 
We construct an optimisation problem to obtain the optimal MDA timings reflecting that our objective is elimination. Since we use the mean-field approximation of the within-host model, here we optimise the deterministic version of the model to avoid the added complexity. 
Though disease elimination may happen when prevalence is at a low level, for \pv, a low number of blood-stage infections may not be sufficient to guarantee elimination due to the re-established infection by the hypnozoite reservoir. 
Therefore, we define elimination to have occurred when there is no infection in the human and mosquito populations, i.e., $I+L+E_m+I_m=0$. 
Formally, the optimisation problem can be written as

\begin{equation*}
\begin{aligned}
\minimise_{x_0, x_1,\ldots, x_{{N_{\text{MDA}}}-1}} \quad & Z\\
\textrm{s.t.} \quad &\ x_1,\ x_2,\ \ldots,\ x_{{N_{\text{MDA}}}-1}>0,\  x_0\ge0 \  \text{and}\ \sum x_i \le t_{max},
\end{aligned}
\label{eqn:optimum_ses}
\end{equation*} 
where 
\begin{equation*}
\begin{aligned}
Z= \int_{s_1}^{t_{max}} \Big(I(t)+L(t)+E_m(t)+I_m(t)\Big) dt
\end{aligned}
\label{eqn:obj}
\end{equation*}
is the objective function and the $x_i,\ i=1,2,\ldots, {N_{\text{MDA}}}-1$ are the intervals between MDA rounds. 
That is, our objective is to minimise the area under the curve as keeping the total infection low for a longer duration (instead of at an instant) to encourage disease extinction. 
When seasonality is not considered, the time of the first MDA, $s_1$, can be considered arbitrary (as long as the dynamics have reached an equilibrium). 
However, when considering seasonality in the mosquito population, the time of the first MDA round is no longer arbitrary, as the dynamics display periodic oscillations around the mean annual prevalence of blood-stage infection. 
We defined $x_0$ to be the interval between the first MDA round and the most recent peak prevalence. 
We minimise the objective function within a six-year period, starting from the first MDA round (e.g., $t_{max} = 6$ years). 

\begin{figure}[]
\centering
  \includegraphics[width=\textwidth]{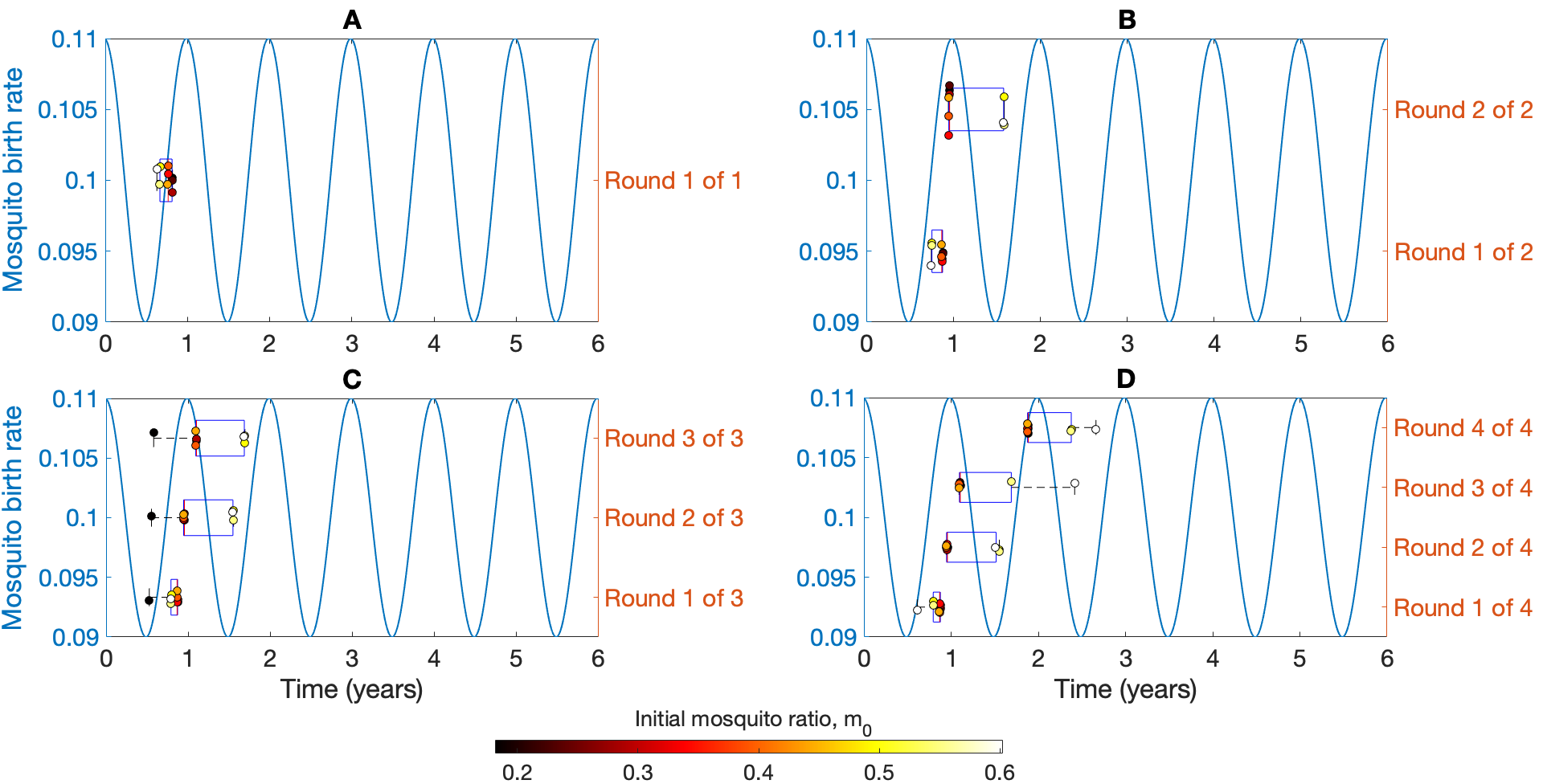}
  \caption{\textit{Optimal MDA times for up to four rounds (right vertical axis) obtained with the deterministic version of the model in reference to seasonal mosquito birth rate $\theta(t)$  (left vertical axis). The box plot shows the distribution of the optimal MDA implementation times for different initial mosquito to human ratios to humans (coloured scattered points). These different initial mosquito ratios correspond to prevalence of blood-stage infection in the range 10--90\%.}}
  \label{fig:MDA_mosq}
\end{figure}

Figure \ref{fig:MDA_mosq} summarises the optimal timing of each of the MDA rounds (up to 4 rounds) for varying initial mosquito to human ratios $m_0$ (hence prevalence), under the model with seasonal dynamics. This figure is produced with the deterministic version of the stochastic model as in Anwar \etal\cite{anwar2023optimal} and shows that the optimal timing of the MDA does not vary considerably as a function of the initial mosquito to human ratio, $m_0$, especially when one (Figure \ref{fig:MDA_mosq}A) and two (Figure \ref{fig:MDA_mosq}B) MDA rounds are under consideration. However, there is variation in optimal MDA implementation times for three and four MDA rounds. For a higher initial mosquito ratio, $m_0$, the optimal interval between MDA rounds is longer (Figures \ref{fig:MDA_mosq}C and D). Note that, with superinfection and seasonality in the mosquito population, a small number of initial mosquitoes can sustain a greater disease prevalence. We utilise these deterministic-model optimal timings in the stochastic model for up to four MDA rounds to study the impact of MDA with radical cure on \pv~elimination. We also investigate another strategy referred to as ``simplified MDA time'', where we implement the first MDA round at a fixed time (after 5 years of burn-in). For this simplified approach the MDA rounds are implemented at 30-day intervals instead of the optimal. The optimal interval between the first and second rounds ($x_1$) when up to four rounds are under consideration, stays close to 29 days when prevalence is 20-60\% (see Table S1 and Figure S2 in Supplementary Material). Hence, we chose an interval of 30 days between all the rounds for the simplified approach. We compare the probability of elimination under this simplified strategy with the optimal strategies.  

\subsection{Model implementation}
To efficiently simulate (approximate) trajectories from the stochastic model (see Algorithm \ref{gillespie}) we use \(\tau\)-leaping \cite{gillespie2001approximate}.
Here, we use the mean-field approximation of the within-host model to obtain the time-dependent parameters $p(t),\ p_1(t),\ p_2(t),\ k_1(t)$, and $k_T(t)$ which makes computation faster.
The initial conditions, including mosquito to human ratio were chosen to have a prevalence of blood-stage infection in the range of 10--90\%, chosen to enable exploration of a wide range of prevalences. We used the MATLAB optimisation tool ‘Multistart’ with 50 different initial starting points in the range [1,365] and with \emph{fmincon} (SQP algorithm) to generate global optimal solutions.
All model parameters are described in Table~\ref{tab:white}.
The optimisation model is implemented to start at year 5 to allow some burn-in period (we obtained the value of $m_0$ using the deterministic model to obtain the equilibrium mean prevalence for the stochastic model).

\begin{algorithm}[!ht]
\caption{Algorithm to approximate an exact trajectory of the stochastic model.}\label{gillespie} 
     Define parameters. Choose $\tau=1$. Initialise model with initial condition $\{S(0), I(0), L(0), S_m(0), E_m(0), I_m(0)\}$.\\
    \While {$t<t_{end}\ \text{or}\ I+L+E_m+I_m=0,$}{
     Calculate the time-dependent parameters $p(t),\ p_1(t),\ p_2(t),\ k_1(t)$, and $k_T(t)$ using Equations (\ref{eqn:p}), \eqref{eqn:p_1}, \eqref{eqn:p_2}, \eqref{eqn:k1}, (\ref{eqn:kT}), respectively.\\
     Calculate all the event rates $R_j,\ j=1,2,\ldots,15$ as in Table~\ref{tab:Stoch}.\\
      % \State Choose time step $\tau=1$.\\
     For each event $E_j,\ j=1,2,\ldots,12$ in Table~\ref{tab:Stoch}, generate $K_j\sim \text{Poisson} (R_j\tau)$ that provides the number of each event that occurs within the time interval $[t,\ t+\tau]$.\\
     Update the states $\{S, I, L, S_m, E_m, I_m\}$ at time $t+\tau$.
      }
\end{algorithm}
 
% \begin{algorithm}[!ht]
% \caption{Algorithm to approximate an exact trajectory of the stochastic model.}
%  \label{gillespie} 
%     \State Define parameters. Choose $\tau=1$. Initialise model with initial condition $\{S(0), I(0), L(0), S_m(0), E_m(0), I_m(0)\}$.\\
%     \While {$t<t_{end}\ \text{or}\ I+L+E_m+I_m=0,$}
%     {
%     \State Calculate the time-dependent parameters $p(t),\ p_1(t),\ p_2(t),\ k_1(t)$, and $k_T(t)$ using Equations (\ref{eqn:p}), \eqref{eqn:p_1}, \eqref{eqn:p_2}, \eqref{eqn:k1}, (\ref{eqn:kT}), respectively.\\
%     \State Calculate all the event rates $R_j,\ j=1,2,\ldots,15$ as in Table~\ref{tab:Stoch}.\\
%       % \State Choose time step $\tau=1$.\\
%       \State For each event $E_j,\ j=1,2,\ldots,12$ in Table~\ref{tab:Stoch}, generate $K_j\sim \text{Poisson} (R_j\tau)$ that provides the number of each event that occurs within the time interval $[t,\ t+\tau]$.\\
%       \State Update the states $\{S, I, L, S_m, E_m, I_m\}$ at time $t+\tau$.
%       }
% \end{algorithm}

\section{Results} \label{result}
\begin{figure}[!ht]
\centering
  \includegraphics[width=.8\textwidth]{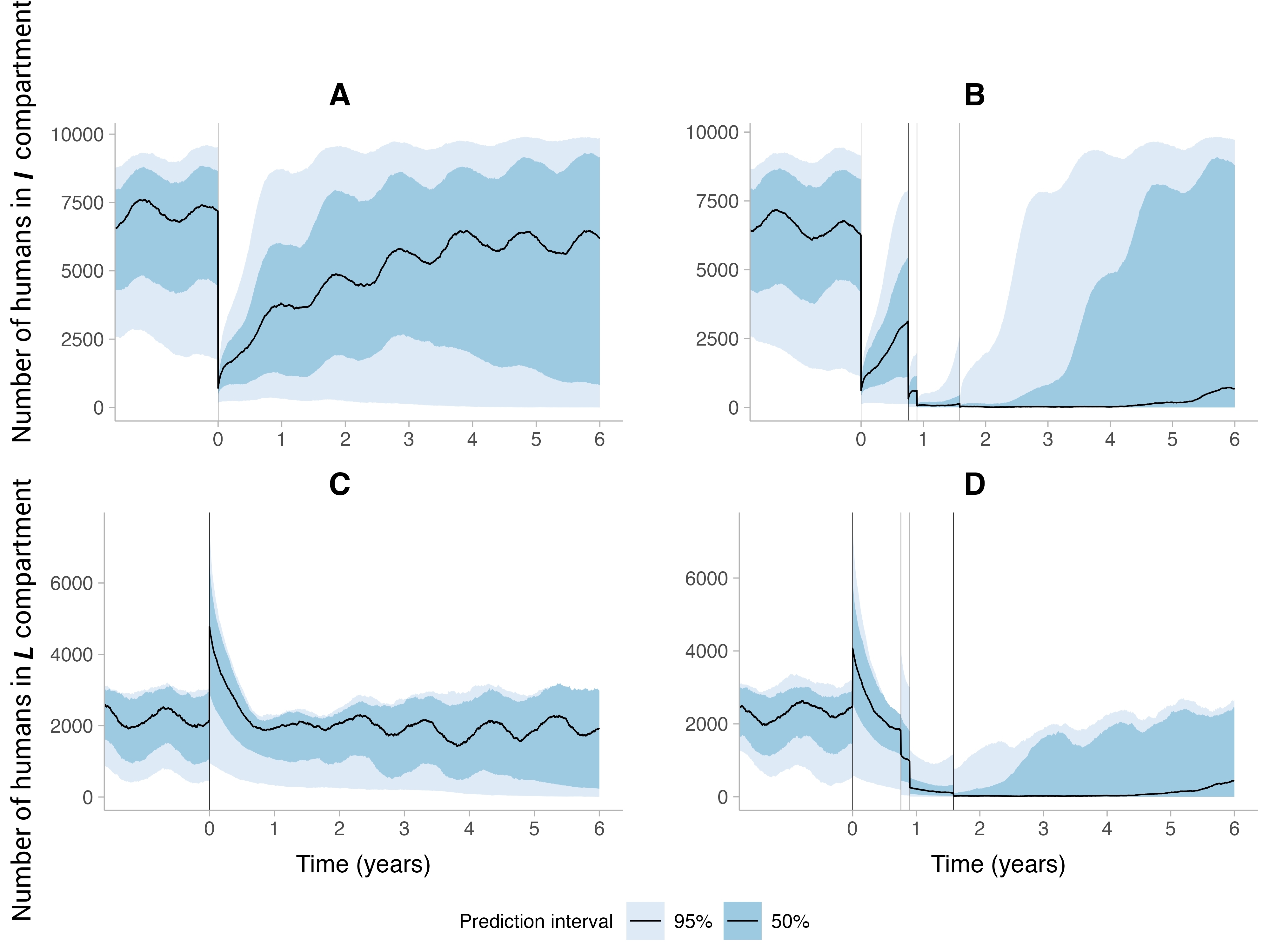}
   \caption{\textit{Effect of one (first column) and four (second column) MDA rounds from 1000 simulations with $T_h=10000,\ m_0=0.38$. Subplots A and C illustrate the effect of one MDA round on the number of humans in the $I$ and $L$ compartments, respectively. Subplots B and D illustrate the effect of four MDA rounds, respectively. The median trajectory out of the 1000 trajectories is indicated by the black line with 50\% and 95\% prediction intervals shown in shading. The grey vertical lines indicate the time of each MDA round. Parameters are as in Table~\ref{tab:white}.}}
  \label{fig:MDA1_MDA4}
\end{figure}

We consider up to four rounds of MDA to study the impact of radical cure treatment on \pv~elimination. 
Figure \ref{fig:MDA1_MDA4} presents the impact of one and four rounds of MDA on the number of humans in the $I$ and $L$ compartments. Here, the timings of the MDA rounds (vertical lines) are chosen to be the optimal MDA times from the deterministic model when the initial mosquito to human ratio is $m_0=0.38$ ($m_0$ is chosen here to have a moderate transmission intensity (70\%)). Figures \ref{fig:MDA1_MDA4}A and C depict the effect of one MDA round, while Figures \ref{fig:MDA1_MDA4}B and D depict the effect of four MDA rounds on the number of humans in the $I$ and $L$ compartments, respectively. As we assume that the efficacy of the radical cure drug is 90\%, the hypnozoite reservoir is never fully cleared when the drug is administered. Therefore, many individuals will transition to the $L$ compartment, explaining the spikes in Figures \ref{fig:MDA1_MDA4}C and D when the first round of MDA is administered. Under four MDA rounds, the median trajectories for both infections reach close to zero (Figures \ref{fig:MDA1_MDA4}B, D) after the fourth round. However, new mosquito bites and hypnozoites from new bites, as well as those hypnozoites that survived the radical cure (because of imperfect drug efficacy), contribute to new infections. This drives the median trajectory for the number of blood-stage infections to eventually increase (Figures \ref{fig:MDA1_MDA4}A, B), albeit with high levels of uncertainty.\\

 \begin{figure}[!ht]
\centering
  \includegraphics[width=\textwidth]{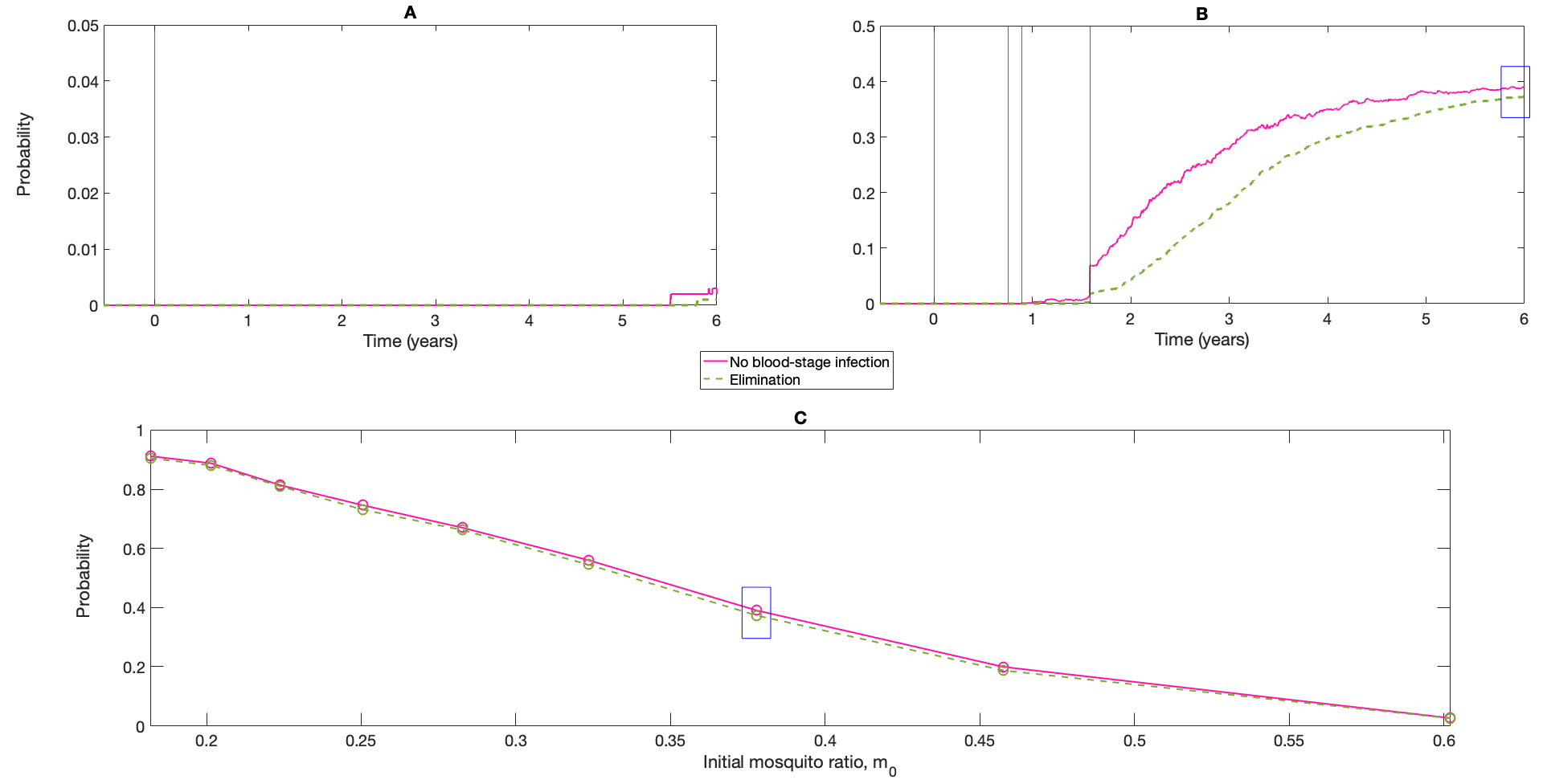}
   \caption{\textit{Probability of \pv~elimination (green) and probability of no blood-stage infection (magenta) over time under one (Subplot A) and four (Subplot B) MDA rounds, respectively, with the initial mosquito to human ratio, $m_0=0.38$. The vertical lines indicate the MDA implementation times. Subplot C: Probability of \pv~elimination and probability of no blood-stage infection for a varying number of initial mosquito to human ratios, $m_0$, under four MDA rounds. Probabilities are evaluated after six years from the first MDA round (from 1000 stochastic simulations). The blue box in Subplot C illustrates the evaluated probability in Subplot B. Note the different scales of the vertical axes in the subplots. Parameters are as in Table~\ref{tab:white}.}}
  \label{fig:Prb_eli_time}
\end{figure}

 \begin{figure}[!ht]
\centering
  \includegraphics[width=\textwidth]{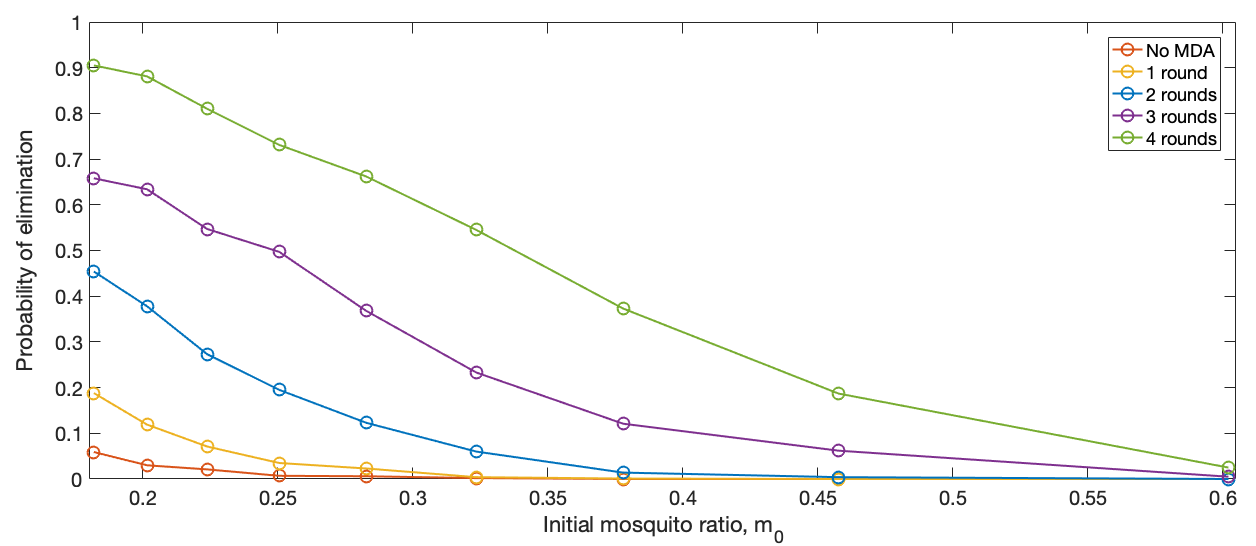}
   \caption{\textit{Effect of up to four rounds of MDA on the probability of \pv~elimination after six years with varying number of initial mosquito to human ratios, $m_0$, compared to no MDA (red line) from 1000 stochastic simulations. Parameters are as in Table~\ref{tab:white}.}}
  \label{fig:Prb_eli_all}
\end{figure}

\pv~ transmission dynamics are primarily dominated by hypnozoite dynamics as an estimated 79--96\% of all \textit{P. vivax} infections are due to relapse (in the absence of radical cure treatment) \cite{luxemburger1999treatment,baird2008real, betuela2012relapses,commons2018effect,commons2019risk}. Hence, even if the number of blood-stage infections reaches zero, the activation of hypnozoites can contribute to new blood-stage infections, re-initiating or sustaining transmission. Figure \ref{fig:Prb_eli_time} depicts the probability of elimination (green) and the probability of no blood-stage infection (magenta) over time. Figures \ref{fig:Prb_eli_time}A and B depict the probability under one and four MDA rounds, respectively, when the initial mosquito to human ratio is $m_0=0.38$. The probability of elimination and the probability of no blood-stage infection under one MDA are very small (Figure \ref{fig:Prb_eli_time}A). That is, one MDA round does not have any epidemiologically relevant effect on the probability of elimination when the transmission intensity is moderate to high (for the assumed human population size, $T_h=10000$ and initial mosquito population size, $T_m=3800$). However, the probability of elimination and the probability of no blood-stage infection under four MDA rounds increases over time after the fourth MDA round and is epidemiological relevant (in terms of magnitude), see Figure \ref{fig:Prb_eli_time}B. Since we assume that any ongoing blood-stage infection will be cleared instantaneously due to the efficacy of the drug (here $p_{\text{blood}}=0.9$), 
the number of blood-stage infections is driven down to very low numbers immediately following the fourth MDA round (as per Figures \ref{fig:MDA1_MDA4}B).
 The stochastic event of blood-stage elimination is more likely when blood-stage infection numbers are low, which explains why the probability of no blood-stage infection increases after the fourth MDA round. 
 Furthermore, since we assume that hypnozoites in the liver are cleared instantaneously upon administration of the radical cure, the effect of radical cure on hypnozoites not only depends on the efficacy of the drug (here each hypnozoite is cleared with probability $p_{\text{rad}}=0.9$) but also on the hypnozoite reservoir size ($k_T$). 
Activation of the hypnozoites that survive radical cure, and subsequent infectious mosquito bites, contribute to new blood-stage infections; thus the probability of no blood-stage infection over time is not monotonically increasing over time. 
This is distinct from the probability of elimination, which monotonically increases since elimination is an absorbing state of the system. The probability of elimination and the probability of no blood-stage infection vary across different transmission intensities. 
Figure \ref{fig:Prb_eli_time}C depicts the probability of elimination and the probability of no blood-stage infection under four rounds of MDA for varying initial mosquito to human ratios, $m_0$, six years after the first MDA round. The lower the initial mosquito to human ratio, $m_0$, the higher the probabilities of elimination and no blood-stage infection. We note that these results are for a fixed human population size, $T_h=10000$; the probability of elimination would vary with $T_h$.\\

The probabilities of \pv~elimination after six years for up to four rounds of MDA, and no MDA, for varying initial mosquito to human ratios, $m_0$, are depicted in Figure \ref{fig:Prb_eli_all}. The overall impact of one MDA round (yellow line, Figure \ref{fig:Prb_eli_all}) is not substantially different compared to no MDA (red line, Figure \ref{fig:Prb_eli_all}), particularly as the initial mosquito ratio ($m_0$) increases. When the initial mosquito to human ratio, $m_0$, is low, a small proportion of simulations fade out even if there are no rounds of MDA, which corresponds to a low probability of elimination. However, as the transmission intensity increases with increasing $m_0$, there is negligible probability of elimination without MDA. As $m_0$ increases, the probability of elimination decreases (regardless of MDA number and timing). Furthermore, the probability of elimination increases for fixed $m_0$ as the number of MDA rounds increases. That is, with a higher number of MDA rounds (and $m_0$ held fixed), the probability of \pv~elimination is higher.\\

 \begin{figure}[!ht]
\centering
  \includegraphics[width=\textwidth]{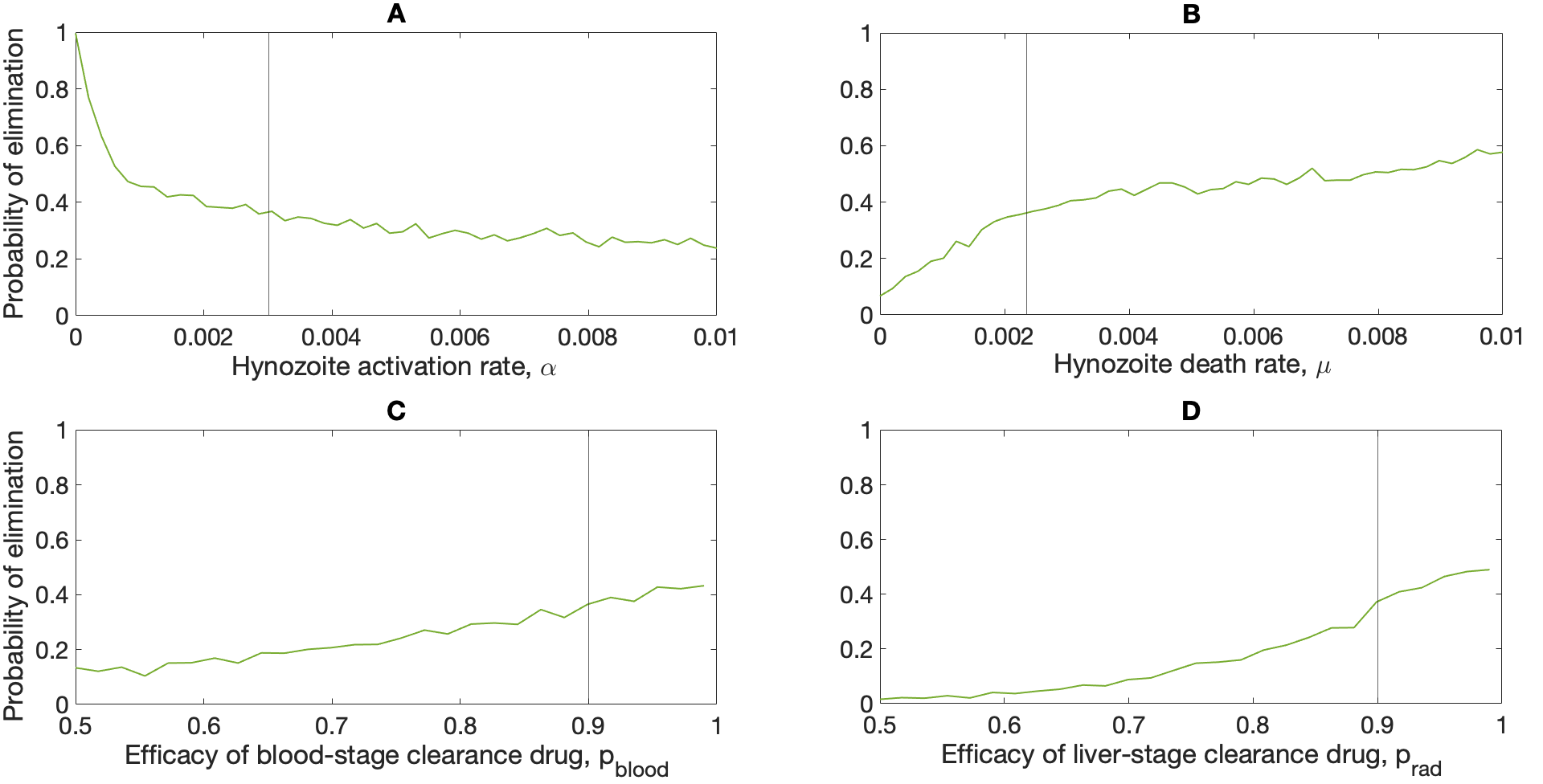}
   \caption{\textit{Sensitivity of probability of \pv~elimination after six years over key model parameters under four rounds of MDA from 1000 stochastic simulations. Subplots A--D depict the probability of elimination over hypnozoite activation rate, $\alpha$, hypnozoite death rate, $\mu$, the efficacy of blood-stage clearance drug, $p_{\text{blood}}$, and the efficacy of liver-stage clearance drug, $p_{\text{rad}}$, respectively. The vertical line in each subplot indicates the baseline parameters used throughout Figures \ref{fig:MDA_mosq}~--\ref{fig:Prb_eli_all}. Probabilities are evaluated after six years from the first MDA round. All other  parameters are as in Table~\ref{tab:white}.}}
  \label{fig:sen_eli}
\end{figure}

Hypnozoite dynamics and the efficacy of radical cure have a considerable influence on \pv~transmission and elimination. Figure \ref{fig:sen_eli} illustrates the sensitivity of the probability of \pv~elimination to some key model parameters under four rounds of MDA for a fixed $m_0=0.38$.  The initial mosquito to
human ratio $m_0$ is chosen here to give a moderate transmission intensity and a moderate chance of \pv~elimination after six years (see Figure \ref{fig:Prb_eli_time}C). Note that the timing of the four MDA rounds is derived optimally (using the deterministic version of the model) for the baseline set of parameter values (vertical line in each subplot in Figure \ref{fig:sen_eli}) from Table~\ref{tab:white}. 
The residual variation evident in each subplot is due to the finite sample size (here 1000 model simulations were considered) used to compute Monte Carlo estimates for the probability of elimination. Figure \ref{fig:sen_eli}A depicts the probability of elimination over the hypnozoite activation rate, $\alpha$, where the vertical line indicates the baseline rate as per Table~\ref{tab:white}. When the activation rate, $\alpha$, is zero, the only option for the hypnozoite is to die. In such cases, infectious mosquitoes are the only driver of ongoing disease transmission. Hence, with four rounds of MDA, there is a high chance of eliminating \pv~with a probability very close to one. However, as we have assumed that radical cure does not provide protection from new infections, the disease may re-establish. As the activation rate, $\alpha$, increases from zero, relapse in addition to infectious bites contributes to onward transmission, and the probability of elimination decreases. 

The probability of elimination as a function of the hypnozoite death rate, $\mu$, under four rounds of MDA is depicted in Figure \ref{fig:sen_eli}B. If the hypnozoite death rate, $\mu$, is zero, the hypnozoites can only activate, increasing the disease burden, resulting in a very low probability of elimination (due to stochasticity, some simulations still fade out). The probability of elimination increases as the hypnozoite death rate increases. That is, the higher the death rate, $\mu$, the higher the chance of elimination, as with hypnozoites dying more frequently, the disease burden from relapse decreases. Figures \ref{fig:sen_eli}C--D depict the elimination probability for varying efficacy of the blood-stage and liver-stage drugs, respectively, when the other underlying parameters are as in Table~\ref{tab:white}. Unsurprisingly the probability of elimination increases with the efficacy of the blood-stage clearance drug and is the highest when 100\% effective. The reason that the probability of elimination is not 100\% when the blood-stage drug is 100\% effective is that the effectiveness of the liver-stage clearance drug, $p_{\text{rad}}$, is set at the baseline value of 90\%. Since it is possible that hypnozoites are not fully cleared, in addition to subsequent infectious mosquito bites, relapses can re-establish transmission. 
In the case of liver-stage drugs, the probability of elimination also increases with the efficacy of the liver-stage clearance drug, $p_{\text{rad}}$. Again, the more effective the drugs are, the higher the probability of elimination, peaking when 100\% effective. It is worth noting that, radical cure with a low-efficacy drug, especially the liver-stage clearance drug, has little to no effect. That is, the chances of elimination improve with high-efficacy drugs.

 \begin{figure}[!ht]
\centering
  \includegraphics[width=\textwidth]{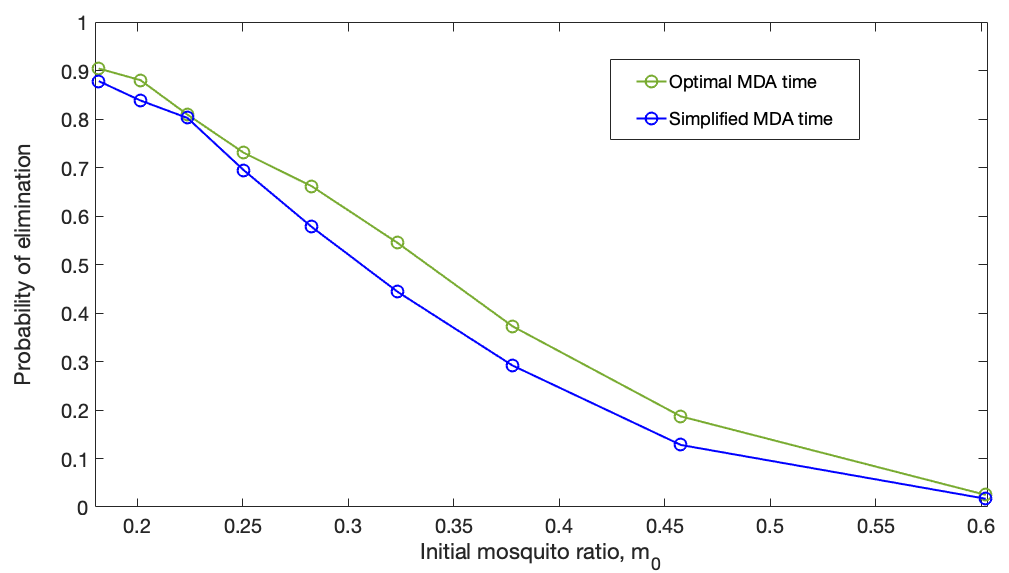}
   \caption{\textit{Probability of \pv~elimination under four rounds of MDA with optimal timing (green line, timings of MDA decided as per the deterministic version of the model) compared to simplified MDA timing (blue solid line) for varying $m_0$ with 1000 simulations. Probabilities are evaluated at year six from the first MDA round. All other parameters are as in Table~\ref{tab:white}.}}
  \label{fig:opt_prac}
\end{figure}
We explored how much the specific timing for MDA implementation affects the probability of \pv~elimination. To illustrate this, we use a simplified way of implementing four MDA rounds. Namely, we implement the MDA rounds with a 30-day interval instead of the optimal intervals (as determined using a deterministic analysis). We note that the choice of this 30-day interval is influenced by the optimal intervals as discussed in Section \ref{obj}. Figure \ref{fig:opt_prac} illustrates the probability of elimination under four MDA rounds with the optimal timing (green line) and compares to the simplified timing (blue line) for varying values of the initial mosquito to human ratio, $m_0$. Here, for every choice of the initial mosquito to human ratio, $m_0$, the optimal approach (as determined using the deterministic model) provides better results in terms of yielding a higher probability of elimination. However, the differences are not substantial and depending on other potential factors, policymakers could also consider a simplified strategy based on resources. 

\section{Discussion} \label{discussion}
As \pv~transmission is primarily dominated by hypnozoite dynamics, targeting the hypnozoites with a radical cure to disrupt transmission is crucial to achieve the goal of \pv~elimination \cite{anwar2023optimal,nekkab2023accelerating}. In this article, we study the impact of multiple rounds of MDA with radical cure on \pv~elimination, defined to be when there are no infections in both human and mosquito populations. 
We previously studied the optimal interruption of \pv~transmission with MDA under a deterministic multiscale model (see Anwar \etal\cite{anwar2023optimal}), but since this model was developed in a deterministic framework, disease elimination was not able to be considered. 
Here we have implemented a stochastic model to allow us to investigate elimination. 
We evaluate the probability of \pv~elimination after six years from the first MDA, which has been chosen to fit with the current public health goals to achieve malaria elimination in at least 35 countries by 2030 according to WHO's Global Technical Strategy \cite{mberikunashe2021onyango}. 
We observed that, as expected, the more MDA rounds there are, the better the chance of \pv~elimination (Figure \ref{fig:Prb_eli_all}).
The administration times of the MDA rounds, when multiple MDA rounds are under consideration can affect the probability of elimination (Figure \ref{fig:opt_prac}). Our results demonstrate that reductions in transmission due to MDA would likely be short-lived (unless elimination is achieved) as the underlying driver of transmission remains unchanged by the intervention.
As the abundance of mosquitoes greatly affects  \pv~dynamics, the higher the mosquito population sizes are, the lower the probability of elimination, regardless of how many MDA rounds are considered, up to the four we explored (Figure \ref{fig:Prb_eli_all}). This is also true for the human population size (not explicitly shown here).  We also observed that it is important to consider a radical cure with high efficacy (Figure \ref{fig:sen_eli}D). Otherwise, the impact of MDA rounds on \pv~elimination probability could be limited (this also depends on other model parameters). However, even with highly effective drugs and varying numbers of MDA rounds, elimination is anticipated to be unlikely with MDA alone (for example, the probability of elimination does not reach 100\% for any parameter combinations considered in Figure \ref{fig:sen_eli}). 

The possibility of \pv~ elimination under MDA depends on many different factors, for example, how effective the drugs are and what proportion of the population is covered under MDA \cite{asih2018challenges,thriemer2021towards}. In this work, we assumed that radical cure drugs are 90\% effective in clearing both blood-stage parasites and hypnozoites, which falls in the estimated efficacy region of 57·7--95\% (see Figures \ref{fig:sen_eli}C--D) depending on the combination of drugs \cite{huber2021radical,nelwan2015randomized,llanos2014tafenoquine}. Since we explicitly model the impact of radical cure drugs on both blood-stage infections and hypnozoite dynamics, for simplicity, we assumed 100\% treatment coverage in our model, which is difficult to achieve in reality due to various factors \cite{agboraw2021factors,finn2020treatment}. Furthermore, human and mosquito movement in and out of the study area may also result in the potential reintroduction of infection, which we did not consider \cite{das2023modelling}, though reactivation still results in a type of reintroduction into the target population. 
This means that, in regards to this parameter, our results are optimistic: our model's estimate of the probability of elimination is an upper bound. Furthermore, because of the risk of hemolysis in G6PD–deficient individuals, radical cure is not recommended by the WHO without screening for G6PD deficiency \cite{world2021second,howes2012g6pd, watson2018implications}. Currently, We do not consider G6PD deficiency in our model. Therefore, accounting for G6PD deficiency is a potential avenue for future work and accounting for it will reduce the probability of elimination since fewer people will be able to take radical cure drugs.

Furthermore, because of the extensive use of antimalarial drugs, the parasite has developed resistance to some drugs, particularly chloroquine which is another reason MDA is not recommended. However, chloroquine is still effective in most parts of the world for \textit{P. vivax} \cite{world20world}. We currently do not consider drug resistance in our model. Moreover, the role of immunity can greatly influence how malaria dynamics progress through the population. However, we do not consider immunity. Individuals who have not previously experienced malaria infection almost invariably become infected when first exposed to infectious mosquito bites, as immunity against malaria has not yet developed \cite{langhorne2008immunity}. Repeated exposure to infectious bites will still likely result in infection, though these individuals may be protected against severe malaria or death \cite{langhorne2008immunity}. However, the immunity acquired from a primary infection may protect more strongly against relapses (which are genetically related to the primary infection) than against a new, genetically distinct primary infection \cite{anwar2023scoping}. 
This is because the parasites could be genetically identical or related, which could elicit a more protective immune response due to familiarity with the primary infection \cite{white2011determinants,joyner2019humoral}. Thus, relapses from the same batch of hypnozoites may be more likely only to cause asymptomatic infections. 
Since the multiscale model presented here depends on the history of past infections, including the role of immunity is, therefore, a potential future work as immunity against \pv~ strongly correlates to past infections. 

From a methodological perspective, another limitation of our work is that we utilised the optimal MDA implementation times from a deterministic analysis of the stochastic version of the model. 
We also have not calibrated our model to any data. Therefore, all these are fruitful avenues for future work.  

Mosquito dynamics greatly influence \pv~dynamics, and as such, vector interventions, for example, long-lasting insecticide nets (LLIN) and indoor residual spray (IRS), should be considered along with MDA. Although we do not explicitly consider vector interventions in this paper, in areas where vector interventions are applied consistently,  they can be implicitly considered through $m_0$ by assuming an overall (constant) effect on the mosquito population (see Figure \ref{fig:opt_prac}). We note there have been substantial efforts invested in explicitly exploring mosquito interventions \cite{white2014modelling,white2018mathematical,nekkab2021estimated}.
The primary purpose of this work was to study the impact of MDA on \pv~elimination; hence, studying the impact of MDAs along with other vector interventions is left for future work.
Nonetheless, as our model accounts for stochasticity and hypnozoite dynamics will more detail than other published models, and we model the impact of MDA on each hypnozoite and infection independently, our research has the potential to contribute towards the goal of \pv~elimination.

\section*{Acknowledgement}
This research was supported by The University of Melbourne’s Research Computing Services and the Petascale Campus Initiative.
J.A. Flegg’s research is supported by the Australian Research Council (DP200100747, FT210100034) and the National Health and Medical Research Council (APP2019093). 
J.M. McCaw’s research is supported by the Australian Research Council (DP210101920) and the NHMRC Australian Centre of Research Excellence in Malaria Elimination (ACREME) (APP1134989).
The contents of the published material are solely the responsibility of the individual authors and do not reflect the views of NHMRC. 

\printbibliography[heading=bibintoc]
 % \bibliography{Ref.bib}
 % \subfile{Supplementary_file.tex}

 %%%%%%%%%%%%%%%% SUPPLEMENTARY

\pagebreak

% \onecolumngrid
\begin{center}
  \textbf{\large Investigation of \textit{P. Vivax} Elimination via Mass Drug Administration-\\Supplementary Material}\\[.2cm]
  Md Nurul Anwar$^{1,*}$ James M. McCaw$^{1,2}$ Alexander E. Zarebski $^1$,Roslyn I. Hickson$^{1,3,4,\dag}$ and Jennifer A. Flegg\\[.1cm]
  {\itshape ${}^1$1School of Mathematics and Statistics, The University of Melbourne, Parkville, Australia\\
  ${}^2$2Centre for Epidemiology and Biostatistics, Melbourne School of Population and Global Health,
The University of Melbourne, Parkville, Australia\\
  ${}^3$3Australian Institute of Tropical Health and Medicine, James Cook University, Townsville,
Australia\\
${}^4$4CSIRO, Townsville, Australia\\}
  ${}^*$Corresponding author: nurul.anwar@unimelb.edu.au\\
  ${}^\dag$These authors contributed equally to this work\\
% (Dated: \today)\\[1cm]
\end{center}

\setcounter{equation}{0}
\setcounter{section}{0}
\setcounter{figure}{0}
\setcounter{table}{0}
\setcounter{page}{1}
\renewcommand{\theequation}{S\arabic{equation}}
\renewcommand{\thefigure}{S\arabic{figure}}
\renewcommand{\thetable}{S\arabic{table}}
\renewcommand{\thesection}{S\arabic{section}}
% \newrefcontext[labelprefix=S]
% \renewcommand{\bibnumfmt}[1]{[S#1]}
% \renewcommand{\citenumfont}[1]{S#1}

\section{Within-host model} \label{ch_m/withinhost}

To capture the hypnozoite dynamics and its stochasticity within individuals, we embed a within-host model into our population level model. 
The within-host model was developed by Mehra \textit{et al.} \parencite{mehra2022hypnozoite}.
Our primary purpose in using the within-host model is to evaluate the probabilities $p(t)$, $p_1(t)$, $p_2(t)$, $k_1(t)$ and $k_T(t)$ under multiple MDA rounds.
These quantities describe aspects of the distribution of the hypnozoite reservoir and blood-stage infection within an individual; we use them to specify the dynamics at the population level.

% {\color{purple} AEZ: I think it would be helpful to use a bit more of the terminology of queuing theory here to specify the model. If this is going to a more mathematical audience, I think they will appreciate the formalism. E.g. mosquitoes arrive in the queue following a Poisson process. E.g. there are infinite servers in the queue so each hypnozoite is modelled independently. I'm a bit foggy on Kendall's notation, but I think this means we have a $\text{M}^{X}/\text{PH}/\infty$ queue, i.e. Markov arrivals in batchs of $X\sim\text{Geom}(p)$ with a phase-type distribution for service times and infinite servers.}

Mosquito bites are modelled as a non-homogeneous Poisson process with rate $\lambda(t)$.
Each bite introduces a batch of hypnozoites.
The number of hypnozoites in the batch is geometrically distributed. 
The hypnozoites then independently progress through different states.
This model can be considered as an infinite server queue process ($M_t^X/G/\infty$) where $X\sim\text{Geom}(\nu)$ \cite{mehra2022hypnozoite}. 

\begin{figure}[!ht]
    \begin{centering}
        \includegraphics[width=1\textwidth]{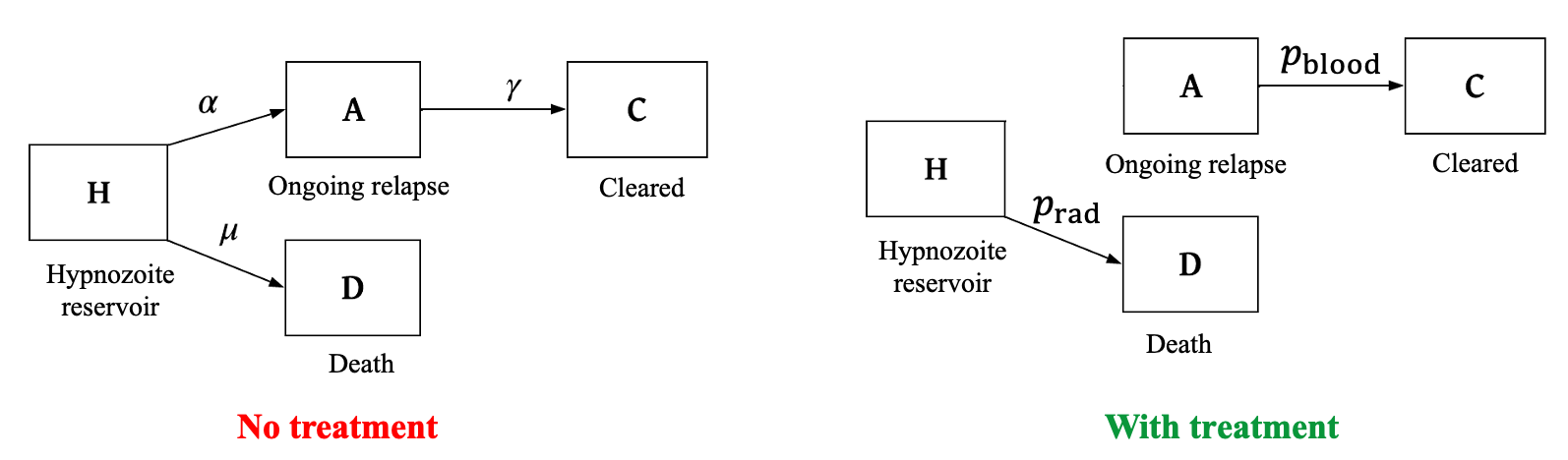}
    \caption{\textit{The within-host model for a single hypnozoite without (left) and with treatment (right). The levels $H$, $A$, $C$, and $D$ represent states of establishment, activation, clearance and death of the hypnozoite, respectively. Parameters $p_{\text{blood}}$ and $p_{\text{rad}}$ account for the instantaneous effect of the treatment.}}
    \label{fig:WH_w_wo_treatment}
    \end{centering}
\end{figure}

Upon arrival within the host, hypnozoites enter a reservoir.
Subsequently, each hypnozoite activates at a constant rate, $\alpha$, (which immediately triggers a blood-stage infection that is cleared at a constant rate, $\gamma$) or dies at a constant rate, $\mu$, due to the death of the host cell.
For the short-latency case (in which hypnozoites can immediately activate after establishment without going through a latency phase), a hypnozoite can be in one of four different states.
Let $H$, $A$, $C$, and $D$ represent states of establishment, activation, clearance (removal after activation) and death (removal before activation) of the hypnozoite, respectively.
Each hypnozoite has two possible final states: death before activation $(D)$; or clearance after activation $(C)$. 
Figure \ref{fig:WH_w_wo_treatment} illustrates the possible trajectories of a single hypnozoite with and without any treatment.

The distribution of the state of a single hypnozoite (in the absence of treatment) is described by the following system of differential equations:

\begin{equation}
\begin{aligned}
    \label{eqn:ch_m/3.1}
    \frac {dp_H}{dt}=&-(\alpha+\mu)p_H,\\
    % \label{eqn:ch_m/3.2}
    \frac {dp_A}{dt}=&\alpha p_H-\gamma p_A,\\
    % \label{eqn:ch_m/3.3}
    \frac {dp_C}{dt}=&\gamma p_A,\\
    % \label{eqn:ch_m/3.4}
    \frac {dp_D}{dt}=&\mu p_H,
\end{aligned}
\end{equation}
\noindent
with the initial condition $\mathrm{\bf p}(0)=(p_H(0),p_A(0),p_C(0),p_D(0))=(1,0,0,0)$.

We can solve the equations to get the following distribution of the hypnozoite state through time:
\begin{equation}
\begin{aligned}
    \label{eqn:ch_m/3.5}
    p_H(t)=&e^{-(\alpha +\mu)t},\\
    % \label{eqn:ch_m/3.6}
    p_A(t)=&\frac{\alpha}{(\alpha+\mu)-\gamma}\left(e^{-\gamma t}-e^{-(\alpha +\mu)t}\right),\\
    % \label{eqn:ch_m/3.7}
    p_C(t)=&\frac{\alpha}{\alpha+\mu}\left(1-e^{-(\alpha +\mu)t}\right)-\frac{\alpha}{(\alpha+\mu)-\gamma}\left(e^{-\gamma t}-e^{-(\alpha +\mu)t}\right),\\
    % \label{eqn:ch_m/3.8}
    p_D(t)=&\frac{\mu}{\alpha+\mu}\left(1-e^{-(\alpha +\mu)t}\right).
\end{aligned}
\end{equation}

\subsection{Treatment model}

% {\color{purple} AEZ: The following derivation does not explicitly consider that the arrival time of the hynozoite is itself a variable, i.e. the zero point. I think this might be implicitly managed in the integrals below through \(p_{f}(t-\tau)\) but I'm not sure. I think this will be a little easier to follow if you add a few comments to spell this out.}

Suppose that treatment is administered at times $s_1 <s_2<\ldots<s_{N_{\text{MDA}}}$. 
Considering a single hypnozoite, we denote its state at time $t$ with
$X_r(t,\ s_1,\ s_2,\ \ldots, s_{N_{\text{MDA}}})\in \{H,A,C,D\}$ with corresponding PMF 
$$p^r_H(t,s_1,\ldots,s_{N_{\text{MDA}}}),\ p^r_A(t,s_1,\ldots,s_{N_{\text{MDA}}}),\ p^r_C(t,s_1,\ldots,s_{N_{\text{MDA}}}),\ p^r_D(t,s_1,\ldots,s_{N_{\text{MDA}}}),$$
where the superscript $``r"$ is for radical cure. Note that, for $t<s_1$, $\mathbf{p}^r(t,s_1,s_2,s_3,\ldots, s_{N_{\text{MDA}}})=\mathbf{p}(t)$ as in Equation (\ref{eqn:ch_m/3.5}).When the treatment is administered, any ongoing relapse will be cleared instantaneously with probability $p_{\text{blood}}$ and each hypnozoite will die instantaneously with probability $p_{\text{rad}}$.
The governing equations for the state probabilities under treatment (as described by Mehra \textit{et al.}\ \cite{mehra2022hypnozoite}) are
\begin{align}
\label{eqn:1}
    \frac {dp^r_H}{dt}=&-(\alpha+\mu)p^r_H-\ln{\big((1-p_{\text{rad}})\big)^{-1}}\sum_{j=1}^{N_\text{MDA}} \delta_D(t-s_j)p^r_H,\\
    \label{eqn:2}
    \frac {dp^r_A}{dt}=&-\gamma p^r_A+\alpha p^r_H-\ln{\big((1-p_{\text{blood}})\big)^{-1}}\sum_{j=1}^{N_\text{MDA}} \delta_D(t-s_j)p^r_A,\\
    \label{eqn:3}
    \frac {dp^r_C}{dt}=&\gamma p^r_A+\ln{\big((1-p_{\text{blood}})\big)^{-1}}\sum_{j=1}^{N_\text{MDA}} \delta_D(t-s_j)p^r_A,\\
    \label{eqn:4}
    \frac {dp^r_D}{dt}=&-\mu p^r_H+\ln{\big((1-p_{\text{rad}})\big)^{-1}}\sum_{j=1}^{N_\text{MDA}} \delta_D(t-s_j)p^r_H,
\end{align}
 where $\delta_D(\cdot)$ is the Dirac delta function.  Due to the instantaneous assumption of the effect of treatment, the state probabilities $\mathbf{p}^r(t,s_1,s_2,s_3,\ldots, s_{N_{\text{MDA}}})$ will exhibit jump discontinuities at $t=s_1,s_2,s_3,\ldots, s_{N_{\text{MDA}}}$.  
 
The framework to account for continuous mosquito inoculation \parencite{mehra2022hypnozoite} assumes that
\begin{itemize}
    \item the dynamics of hypnozoites are independent and identically distributed, with the PMF without treatment given by Equation (\ref{eqn:ch_m/3.5});
    \item infectious mosquito bites follow a non-homogeneous Poisson process with a time-dependent rate, $\lambda(t)$. The mean number of infectious bites in the interval $(0,t]$ is $q(t)$ where
    \begin{align}
    \label{eqn:qt}
        q(t)=\int_0^t \lambda(\tau)d\tau;
    \end{align}
    \item the number of hypnozoites established by each mosquito bite is geometrically-distributed (\parencite{white2014modelling}) with mean $\nu$;
    \item each infectious bite causes a primary infection which is cleared at rate $\gamma$;
    \item and hypnozoites die due to the death of the host liver cell at rate $\mu$ (e.g., there is no administration of anti-hypnozoital drugs).
    % \item individuals are first exposed to infective mosquito bites at time $t=0$.
\end{itemize}

Since we need the distribution of the size of the hypnozoite reservoir and infection status to use in our population-level model, we seek to obtain expressions for the quantities $p(t),\ p_1(t),\ p_2(t),\ k_1(t)$ and $k_T(t)$ from the within-host model under multiple rounds of MDA.
Evaluating the quantities $p(t),\ p_1(t),\ p_2(t),\ k_1(t)$ and $k_T(t)$ in the population-level model requires the probability of hypnozoite establishment ($p^r_H(t)$) and the probability of hypnozoite activation ($p^r_A(t)$) \cite{anwar2023optimal}; hence we solve Equations (\ref{eqn:1})--(\ref{eqn:2}) for $N_{\text{MDA}}$ rounds for $t\ge s_{N_{\text{MDA}}}$ to give
 \begin{align}
 \label{eqn:pH_r}
    p^r_H(t,s_1,s_2,s_3,\ldots, s_{N_{\text{MDA}}})=&(1-p_{\text{rad}})^{N_\text{MDA}} p_H(t),\\
    \label{eqn:pA_r}
    p^r_A(t,s_1,s_2,s_3,\ldots, s_{N_{\text{MDA}}})=&(1-p_{\text{blood}})e^{-\gamma(t-s_{N_{\text{MDA}}})}p_A^r(s_{N_{\text{MDA}}},s_1,s_2,\ldots,s_{N_\text{MDA}-1})\nonumber\\  &+(1-p_{\text{rad}})^{N_\text{MDA}}\big(p_A(t)-e^{-\gamma(t-s_{N_{\text{MDA}}})}p_A(s_{N_{\text{MDA}}})\big),
\end{align}
where $p_H(t)$ and $p_A(t)$ are the probability of establishment and activation of a hypnozoite without treatment, respectively, and are given by Equation \eqref{eqn:3.5}

Let us define two additional states, $P$ and $PC$, to denote an ongoing primary infection from infective mosquito bites and a cleared primary infection, respectively. 
Let $N_f(t)$ denote the number of hypnozoites in states  $f\in F:= \{H,A,C,D\}$ at time $t$ and $N_P(t),\ N_{PC}(t)$ denote the number of ongoing and cleared primary infections, respectively, at time $t$. 
For notational convenience, let $F':=\left\{H,A,C,D,P,PC\right\}$. We can now consider the random vector combining all of these variables:

$$\mathbf{N} (t)=(N_H(t),N_A(t),N_C(t),N_D(t),N_P(t),N_{PC}(t))$$
\noindent
with $\mathbf{N} (0)=\mathbf{0}$. The PGF for $\mathbf{N}$ can be written following from Equation (30) in Mehra \textit{et al.} \parencite{mehra2022hypnozoite} (for short-latency case ($k=0$) with the probability of getting blood-stage infection after an infectious bite, $p_{prim}=1$) as
\begin{align}
    \label{PGF_ntrt}   G(t,&z_H,z_A,z_C,z_D,z_P,z_{PC}):=\mathbb{E}\left[\displaystyle\prod_{f\in F'} z_f^{N_f(t)}\right],\nonumber\\&=\text{exp}\left\{-q(t)+\int_0^t \frac{\lambda(\tau)\left(z_Pe^{-\gamma (t-\tau)}+(1-e^{-\gamma (t-\tau)})z_{PC}\right)}{1+\nu \left(1-\sum_{f\in F} z_fp_f(t-\tau)\right)}d\tau \right\},
    \end{align}
where $q(t)$ is the mean number of infective bites in the interval $(0,t]$ and is given by Equation~(\ref{eqn:qt}).

% {\color{purple} AEZ: Does the RHS of Eq~\eqref{PGF} hold if you allow $t<s_{N_{\text{MDA}}}$? If so, I think that needs to be spelled out, because based on the terminals of the integraion, it looks like it is assuming $t>s_{N_{\text{MDA}}}$. Does there need to be some sort of indicator function for which integrals to include depending on the value of $t$? If not, why? Also, I find it a bit confusing to be referencing the Merha equations so much, if you are using identical notation and it is not too difficult, I think it will be easier to read if you just quote their equations (with references) where they are needed. Or if you could give a few more comments explaining how the resulting expressions were obtained, that might be helpful. There may be similar issues re incomplete $t$ domains in Eqs~\eqref{eqn:nHAP0},\eqref{eqn:nH0}, and \eqref{eqn:nAP0}}

The expression for the joint PGF with drug administration at time $t=s_1$, that is, under one round is given by Equation (31) in Mehra \textit{et al.} \cite{mehra2022hypnozoite}. Following a similar analysis, if the drug is administered at $N_{\text{MDA}}$ successive times ($t=s_1,\ s_2,\ldots, s_{N_{\text{MDA}}}$), then the joint PGF for the number of hypnozoites/infections in each state at time $t$ is
\begin{align}
    &G^{s_1,s_2,\ldots s_{N_{\text{MDA}}}}(t,z_H,z_A,z_C,z_D,z_P,z_{PC}):=\mathbb{E}\left[\displaystyle\prod_{f\in F'} z_f^{N^{s_1,s_2,\ldots s_{N_{\text{MDA}}}}_s(t)}\right], \nonumber\\ 
    &=\begin{cases} 
    \text{exp}\left\{-q(t)+\int_0^t \frac{\lambda(\tau)\left(z_Pe^{-\gamma (t-\tau)}+(1-e^{-\gamma (t-\tau)})z_{PC}\right)}{1+\nu \left(1-\sum_{f\in F} z_fp_f(t-\tau)\right)}d\tau \right\} & \text{if}\ t < s_1 \\ 
    \text{exp}\Bigg\{-q(t)+\int_{s_1}^t \lambda(\tau)\frac{e^{-\gamma (t-\tau)}z_P+(1-e^{-\gamma (t-\tau)})z_{PC}}{1+\nu \left(1-\sum_{f\in F} z_f.p_f(t-\tau)\right)}d\tau  \\ \quad +\int_{0}^{s_1} \lambda(\tau)\frac{(1-p_{\text{blood}})e^{-\gamma (t-\tau)}z_P+(1-(1-p_{\text{blood}})e^{-\gamma (t-\tau)})z_{PC}}{1+\nu \left(1-\sum_{f\in F} z_f.p^r_s(t-\tau,s_1-\tau)\right)}d\tau \Bigg\}& \text{if}\ s_1 \le t< s_2,
    \\
    \text{exp}\Bigg\{-q(t)+\int_{s_2}^t \lambda(\tau)\frac{e^{-\gamma (t-\tau)}z_P+(1-e^{-\gamma (t-\tau)})z_{PC}}{1+\nu \left(1-\sum_{f\in F} z_f.p_f(t-\tau)\right)}d\tau  \\ \quad +\int_{0}^{s_1} \lambda(\tau)\frac{(1-p_{\text{blood}})e^{-\gamma (t-\tau)}z_P+(1-(1-p_{\text{blood}})e^{-\gamma (t-\tau)})z_{PC}}{1+\nu \left(1-\sum_{f\in F} z_f.p^r_s(t-\tau,s_1-\tau)\right)}d\tau \\ \quad +\int_{s_1}^{s_2} \lambda(\tau)\frac{(1-p_{\text{blood}})^2e^{-\gamma (t-\tau)}z_P+(1-(1-p_{\text{blood}})^2e^{-\gamma (t-\tau)})z_{PC}}{1+\nu \left(1-\sum_{f\in F} z_f.p^r_s(t-\tau,s_1-\tau,s_2-\tau)\right)}d\tau \Bigg\} & \text{if}\ s_2 \le t< s_3,
    \\ 
    \quad \vdots
    %  \vdotswithin{}  
    \\
    \text{exp}\Bigg\{-q(t)+\int_{s_{N_{\text{MDA}}}}^t \lambda(\tau)\frac{e^{-\gamma (t-\tau)}z_P+(1-e^{-\gamma (t-\tau)})z_{PC}}{1+\nu \left(1-\sum_{f\in F} z_f.p_f(t-\tau)\right)}d\tau\\ \quad
    +\int_{0}^{s_1} \lambda(\tau)\frac{(1-p_{\text{blood}})e^{-\gamma (t-\tau)}z_P+(1-(1-p_{\text{blood}})e^{-\gamma (t-\tau)})z_{PC}}{1+\nu \left(1-\sum_{f\in F} z_f.p^r_s(t-\tau,s_1-\tau)\right)}d\tau \\ \quad+\int_{s_1}^{s_2} \lambda(\tau)\frac{(1-p_{\text{blood}})^2e^{-\gamma (t-\tau)}z_P+(1-(1-p_{\text{blood}})^2e^{-\gamma (t-\tau)})z_{PC}}{1+\nu \left(1-\sum_{f\in F} z_f.p^r_s(t-\tau,s_1-\tau,s_2-\tau)\right)}d\tau \\ \quad +\int_{s_2}^{s_3} \lambda(\tau)\frac{(1-p_{\text{blood}})^3e^{-\gamma (t-\tau)}z_P+(1-(1-p_{\text{blood}})^3e^{-\gamma (t-\tau)})z_{PC}}{1+\nu \left(1-\sum_{f\in F} z_f.p^r_s(t-\tau,s_1-\tau,s_2-\tau,s_3-\tau)\right)}d\tau \\ \quad+\ldots +\int_{s_{N-1}}^{s_{N_{\text{MDA}}}} \lambda(\tau)\frac{(1-p_{\text{blood}})^{N_{\text{MDA}}}e^{-\gamma (t-\tau)}z_P}{1+\nu \left(1-\sum_{f\in F} z_f.p^r_s(t-\tau,s_1-\tau,s_2-\tau,s_3-\tau,\ldots, s_{N_{\text{MDA}}}-\tau)\right)}+\\
    \qquad \frac{(1-(1-p_{\text{blood}})^{N_{\text{MDA}}}e^{-\gamma (t-\tau)})z_{PC}}{1+\nu \left(1-\sum_{f\in F} z_f.p^r_s(t-\tau,s_1-\tau,s_2-\tau,s_3-\tau,\ldots, s_{N_{\text{MDA}}}-\tau)\right)}d\tau \Bigg\} & \text{if}\ t>s_{N_{\text{MDA}}}.
    \end{cases}
    \label{PGF}
\end{align}

% \begin{align}
%     &G^{s_1,s_2,\ldots s_{N_{\text{MDA}}}}(t,z_H,z_A,z_C,z_D,z_P,z_{PC}):=\mathbb{E}\left[\displaystyle\prod_{f\in F'} z_f^{N^{s_1,s_2,\ldots s_{N_{\text{MDA}}}}_s(t)}\right], \nonumber\\ 
%     &=\text{exp}\Bigg\{-q(t)+\int_{s_{N_{\text{MDA}}}}^t \lambda(\tau)\frac{e^{-\gamma (t-\tau)}z_P+(1-e^{-\gamma (t-\tau)})z_{PC}}{1+\nu \left(1-\sum_{f\in F} z_f.p_f(t-\tau)\right)}d\tau  \nonumber\\ 
%     &  +\int_{0}^{s_1} \lambda(\tau)\frac{(1-p_{\text{blood}})e^{-\gamma (t-\tau)}z_P+(1-(1-p_{\text{blood}})e^{-\gamma (t-\tau)})z_{PC}}{1+\nu \left(1-\sum_{f\in F} z_f.p^r_s(t-\tau,s_1-\tau)\right)}d\tau \nonumber \\& +\int_{s_1}^{s_2} \lambda(\tau)\frac{(1-p_{\text{blood}})^2e^{-\gamma (t-\tau)}z_P+(1-(1-p_{\text{blood}})^2e^{-\gamma (t-\tau)})z_{PC}}{1+\nu \left(1-\sum_{f\in F} z_f.p^r_s(t-\tau,s_1-\tau,s_2-\tau)\right)}d\tau \nonumber\\ & +\int_{s_2}^{s_3} \lambda(\tau)\frac{(1-p_{\text{blood}})^3e^{-\gamma (t-\tau)}z_P+(1-(1-p_{\text{blood}})^3e^{-\gamma (t-\tau)})z_{PC}}{1+\nu \left(1-\sum_{f\in F} z_f.p^r_s(t-\tau,s_1-\tau,s_2-\tau,s_3-\tau)\right)}d\tau \nonumber\\ & +\ldots +\int_{s_{N-1}}^{s_{N_{\text{MDA}}}} \lambda(\tau)\frac{(1-p_{\text{blood}})^{N_{\text{MDA}}}e^{-\gamma (t-\tau)}z_P+(1-(1-p_{\text{blood}})^{N_{\text{MDA}}}e^{-\gamma (t-\tau)})z_{PC}}{1+\nu \left(1-\sum_{f\in F} z_f.p^r_s(t-\tau,s_1-\tau,s_2-\tau,s_3-\tau,\ldots, s_{N_{\text{MDA}}}-\tau)\right)}d\tau \Bigg\}.
%     \label{PGF}
% \end{align}

We now use the PGF in Equations \eqref{PGF_ntrt} and \eqref{PGF} to derive expressions for the probabilities $p(t),\ k_1(t)$, $k_T(t)$, $p_1(t)$, and $p_2(t)$ under multiple MDA rounds.

\subsection{Probability blood-stage infected individual has no hypnozoites:~$p(t)$}
We define $p(t)$ as the probability that an individual has an empty hypnozoite reservoir conditional on an ongoing blood-stage infection (i.e. primary infection or relapse). That is,
\begin{align}
\label{eqn:ch_m/3.12}
  p(t)&=P\big(N_H(t)=0|N_A(t)>0 \cup N_P(t)>0\big)\nonumber\\
  &=\frac{P(N_H(t)=0)-P(N_H(t)=N_A(t)=N_P(t)=0)}{1-P(N_A(t)=N_P(t)=0)},
\end{align}
where the probability that an individual has an empty hypnozoite reservoir at time $t$, $P(N_H(t)=0)$, is given by
\begin{align}
% \label{eqn:3.13}
  P(&N_H(t)=0)=G^{t,s_1,s_2,\ldots,s_{N_{\text{MDA}}}}(t,z_H=0, z_A=1, z_C=1, z_D=1, z_P=1, z_{PC}=1),\nonumber\\
  &=
  \begin{cases}
    \text{exp}\Bigl\{-q(t)+\int_0^t \frac{\lambda(\tau)}{1+\nu p_H(t-\tau)}d\tau \Bigr\}& \text{if}\ t< s_1,
    \\ 
    \text{exp}\Bigl\{-q(t)+\int_{s_1}^t \frac{\lambda(\tau)}{1+\nu p_H(t-\tau)}d\tau+\int_0^{s_1} \frac{\lambda(\tau)}{1+\nu p_H^r(t-\tau,s_1-\tau)}d\tau\\\quad +\int_{s_1}^{s_2} \frac{\lambda(\tau)}{1+\nu p_H^r(t-\tau,s_1-\tau,s_2-\tau)}d\tau \Bigr\}& \text{if}\ s_1 \le t< s_2,
    \\ 
    \quad \vdots  
    \\
    \text{exp}\Bigl\{-q(t)+\int_{s_{N_{\text{MDA}}}}^t \frac{\lambda(\tau)}{1+\nu p_H(t-\tau)}d\tau+\int_0^{s_1} \frac{\lambda(\tau)}{1+\nu p_H^r(t-\tau,s_1-\tau)}d\tau\\\quad +\int_{s_1}^{s_2} \frac{\lambda(\tau)}{1+\nu p_H^r(t-\tau,s_1-\tau,s_2-\tau)}d\tau\\
    \quad+\ldots +\int_{s_{N-1}}^{s_{N_{\text{MDA}}}} \frac{\lambda(\tau)}{1+\nu p_H^r(t-\tau,s_1-\tau,\ldots,s_{N_{\text{MDA}}}-\tau)}d\tau \Bigr\}& \text{if}\ t\ge s_{N_{\text{MDA}}},
    \end{cases} \label{eqn:nH0}
\end{align}

% \begin{align}
% % \label{eqn:3.13}
%   P(&N_H(t)=0)=G^{t,s_1,s_2,\ldots,s_{N_{\text{MDA}}}}(t,z_H=0, z_A=1, z_C=1, z_D=1, z_P=1, z_{PC}=1),\nonumber\\
%   &=
%   \begin{cases}
%     \text{exp}\Bigl\{-q(t)+\int_0^t \frac{\lambda(\tau)}{1+\nu p_H(t-\tau)}d\tau \Bigr\}& \text{if}\ t< s_1,\\ 
%     \text{exp}\Bigl\{-q(t)+\int_{s_{N_{\text{MDA}}}}^t \frac{\lambda(\tau)}{1+\nu p_H(t-\tau)}d\tau+\int_0^{s_1} \frac{\lambda(\tau)}{1+\nu p_H^r(t-\tau,s_1-\tau)}d\tau\\\quad +\int_{s_1}^{s_2} \frac{\lambda(\tau)}{1+\nu p_H^r(t-\tau,s_1-\tau,s_2-\tau)}d\tau\\
%     \quad+\ldots +\int_{s_{N-1}}^{s_{N_{\text{MDA}}}} \frac{\lambda(\tau)}{1+\nu p_H^r(t-\tau,s_1-\tau,\ldots,s_{N_{\text{MDA}}}-\tau)}d\tau \Bigr\}& \text{if}\ t\ge s_{N_{\text{MDA}}},
%     \end{cases} \label{eqn:nH0}
% \end{align}
%
the probability that an individual is neither experiencing a relapse nor a primary infection at time $t$, $P\big(N_A(t)+N_P(t)=0\big)$ (i.e., no blood-stage infection), is given by
\begin{align}
  P&\big(N_A(t)+N_P(t)=0\big)\nonumber\\
  =&G^{t,s_1,s_2,\ldots,s_{N_{\text{MDA}}}}(t,z_H=1, z_A=0, z_C=1, z_D=1, z_P=0, z_{PC}=1), \nonumber \\
  =&
  \begin{cases}
  \text{exp}\Bigl\{-q(t)+\int_0^t \frac{\lambda(\tau)(1-e^{-\gamma (t-\tau)})}{1+\nu p_A(t-\tau)}d\tau\Bigr\} & \text{if}\ t< s_1,
  \\
  \text{exp}\Bigl\{-q(t)+\int_{s_1}^t \frac{\lambda(\tau)(1-e^{-\gamma (t-\tau)})}{1+\nu p_A(t-\tau)}d\tau+\int_0^{s_1} \frac{\lambda(\tau)(1-(1-p_{\text{blood}})e^{-\gamma (t-\tau)})}{1+\nu p_A^r(t-\tau,s_1-\tau)}d\tau\\\quad+\int_{s_1}^{s_2} \frac{\lambda(\tau)(1-(1-p_{\text{blood}})^2e^{-\gamma (t-\tau)})}{1+\nu p_A^r(t-\tau,s_1-\tau,s_2-\tau)}d\tau \Bigl\} & \text{if}\ s_1 \le t< s_2,
  \\ 
    \quad \vdots 
    \\
  \text{exp}\Bigl\{-q(t)+\int_{s_{N_{\text{MDA}}}}^t \frac{\lambda(\tau)(1-e^{-\gamma (t-\tau)})}{1+\nu p_A(t-\tau)}d\tau\\\quad+\int_0^{s_1} \frac{\lambda(\tau)(1-(1-p_{\text{blood}})e^{-\gamma (t-\tau)})}{1+\nu p_A^r(t-\tau,s_1-\tau)}d\tau\\\quad+\int_{s_1}^{s_2} \frac{\lambda(\tau)(1-(1-p_{\text{blood}})^2e^{-\gamma (t-\tau)})}{1+\nu p_A^r(t-\tau,s_1-\tau,s_2-\tau)}d\tau\\\quad+\ldots+\int_{s_{N-1}}^{s_{N_{\text{MDA}}}} \frac{\lambda(\tau)(1-(1-p_{\text{blood}})^{N_{\text{MDA}}}e^{-\gamma (t-\tau)})}{1+\nu p_A^r(t-\tau,s_1-\tau,\ldots,s_{N_{\text{MDA}}}-\tau)}d\tau\Bigr\} & \text{if}\ t\ge s_{N_{\text{MDA}}},
  \end{cases}
  \label{eqn:nAP0}
\end{align}
and
the probability that an individual is neither experiencing an
infection nor has any hypnozoites in their liver at time $t$, $P\big(N_H(t)=N_A(t)=N_P(t)=0\big)$, is given by
\begin{align}
P&\big(N_H(t)=N_A(t)=N_P(t)=0\big)\nonumber\\=&G^{t,s_1,s_2,\ldots,s_{N_{\text{MDA}}}}(t,z_H=0, z_A=0, z_C=1, z_D=1, z_P=0, z_{PC}=1),\nonumber\\
  =&\begin{cases}
  \text{exp}\Bigl\{-q(t)+\int_0^t \frac{\lambda(\tau)(1-e^{-\gamma (t-\tau)})}{1+\nu(p_H(t-\tau)+p_A(t-\tau)}d\tau\Bigr\} & \text{if}\ t< s_1,
  \\
  \text{exp}\Bigl\{-q(t)+\int_{s_1}^t \frac{\lambda(\tau)(1-e^{-\gamma (t-\tau)})}{1+\nu(p_H(t-\tau)+p_A(t-\tau)}d\tau
  \\\quad +\int_0^{s_1} \frac{\lambda(\tau)(1-(1-p_{\text{blood}})e^{-\gamma (t-\tau)})}{1+\nu(p_H^r(t-\tau,s_1-\tau)+p_A^r(t-\tau,s_1-\tau)}d\tau \Bigl\} & \text{if}\ s_1 \le t< s_2,
  \\ 
    \quad \vdots 
    \\
  \text{exp}\Bigl\{-q(t)+\int_{s_{N_{\text{MDA}}}}^t \frac{\lambda(\tau)(1-e^{-\gamma (t-\tau)})}{1+\nu(p_H(t-\tau)+p_A(t-\tau)}d\tau
  \\\quad +\int_0^{s_1} \frac{\lambda(\tau)(1-(1-p_{\text{blood}})e^{-\gamma (t-\tau)})}{1+\nu(p_H^r(t-\tau,s_1-\tau)+p_A^r(t-\tau,s_1-\tau)}d\tau\\\quad +\int_{s_1}^{s_2} \frac{\lambda(\tau)(1-(1-p_{\text{blood}})^2e^{-\gamma (t-\tau)})}{1+\nu(p_H^r(t-\tau ,s_1-\tau,s_2-\tau)+p_A^r(t-\tau ,s_1-\tau,s_2-\tau)}d\tau \\\quad +\ldots+\int_{s_{N-1}}^{s_{N_{\text{MDA}}}} \frac{\lambda(\tau)(1-(1-p_{\text{blood}})^{N_{\text{MDA}}}e^{-\gamma (t-\tau)})}{1+\nu(p_H^r(t-\tau ,s_1-\tau,\ldots ,s_{N_{\text{MDA}}}-\tau)+p_A^r(t-\tau ,s_1-\tau,\ldots ,s_{N_{\text{MDA}}}-\tau)}d\tau\Bigr\}  & \text{if}\ t\ge s_{N_{\text{MDA}}}.
  \end{cases}
  \label{eqn:nHAP0}
\end{align}

\subsection{Probability liver-stage infected individual has one hypnozoite in liver:~$k_1(t)$}\label{sec:i_hyp}

The probability that a liver-stage infected individual has one hypnozoite in the liver at time $t$ (that is, the conditional probability for $N_H(t)$ given an individual
does not have an ongoing blood-stage infection at time $t$) under $N$ MDA rounds is
\begin{align}
% \label{eqn:k1}
k_1(t)=&P(N_H(t)=1|N_A(t)=N_P(t)=0,N_H(t)>0),\nonumber\\
 =&\frac{P(N_H(t)=1|N_A(t)=N_p(t)=0)}{1-P(N_H(t)=0|N_A(t)=N_P(t)=0)},\nonumber\\
 \label{eqn:k1}
 =&\frac{\text{exp}\left\{g(0,t)-g(1,t)\right\}}{1-P(N_H(t)=0|N_A(t)=N_P(t)=0)} \frac{\partial g(0,t)}{\partial z},\nonumber\\
 =&\frac{P(N_H(t)=N_A(t)=N_P(t)=0)}{\big(1-P(N_H(t)=0|N_A(t)=N_P(t)=0)\big)P(N_A(t)=N_P(t)=0)} \frac{\partial g(0,t)}{\partial z},
 \end{align}
where 
\begin{align}
    \frac{\partial g(0,t)}{\partial z}=&\begin{cases}
    \int_0^t \frac{\lambda(\tau)\nu p_H(t-\tau)(1-e^{-\gamma (t-\tau)})}{[1+\nu(p_H(t-\tau)+p_A(t-\tau))]^2}d\tau & \text{if}\ t< s_1,
    \\
    \int_{s_1}^t \frac{\lambda(\tau)\nu p_H(t-\tau)(1-e^{-\gamma (t-\tau)})}{[1+\nu(p_H(t-\tau)+p_A(t-\tau))]^2}d\tau \\ \quad+\int_0^{s_1} \frac{\lambda(\tau)\nu p_H^r(t-\tau,s_1-\tau)(1-(1-p_{\text{blood}})e^{-\gamma (t-\tau)})}{[1+\nu(p_H^r(t-\tau,s_1-\tau)+p_A^r(t-\tau,s_1-\tau))]^2}d\tau\\ \quad +\int_{s_{1}}^{s_2} \frac{\lambda(\tau)\nu p_H^r(t-\tau,s_1-\tau,s_2-\tau)(1-(1-p_{\text{blood}})^2e^{-\gamma (t-\tau)})}{[1+\nu(p_H^r(t-\tau,s_1-\tau,s_2-\tau)+p_A^r(t-\tau,s_1-\tau,s_2-\tau))]^2}d\tau& \text{if}\ s_1 \le t< s_2,
  \\ 
    \quad \vdots 
    \\
    \int_{s_{N_{\text{MDA}}}}^t \frac{\lambda(\tau)\nu p_H(t-\tau)(1-e^{-\gamma (t-\tau)})}{[1+\nu(p_H(t-\tau)+p_A(t-\tau))]^2}d\tau \\ \quad+\int_0^{s_1} \frac{\lambda(\tau)\nu p_H^r(t-\tau,s_1-\tau)(1-(1-p_{\text{blood}})e^{-\gamma (t-\tau)})}{[1+\nu(p_H^r(t-\tau,s_1-\tau)+p_A^r(t-\tau,s_1-\tau))]^2}d\tau\\ \quad+\int_{s_{1}}^{s_2} \frac{\lambda(\tau)\nu p_H^r(t-\tau,s_1-\tau,s_2-\tau)(1-(1-p_{\text{blood}})^2e^{-\gamma (t-\tau)})}{[1+\nu(p_H^r(t-\tau,s_1-\tau,s_2-\tau)+p_A^r(t-\tau,s_1-\tau,s_2-\tau))]^2}d\tau\\ \quad+\ldots
    +\int_{s_{N-1}}^{s_{N_{\text{MDA}}}} \frac{\lambda(\tau)\nu p_H^r(t-\tau,s_1-\tau,\ldots,s_{N_{\text{MDA}}}-\tau)(1-(1-p_{\text{blood}})^{N_{\text{MDA}}}e^{-\gamma (t-\tau)})}{[1+\nu(p_H^r(t-\tau,s_1-\tau,\ldots,s_{N_{\text{MDA}}}-\tau)+p_A^r(t-\tau,s_1-\tau,\ldots,s_{N_{\text{MDA}}}-\tau))]^2}d\tau & \text{if}\ t\ge s_{N_{\text{MDA}}}.
    \end{cases}
\end{align}

Here, the expression
\begin{align*}
    P(N_H(t)=1|N_A(t)=N_P(t)=0)=\text{exp}\left\{g(0,t)-g(1,t)\right\}\frac{\partial g(0,t)}{\partial z}
\end{align*}
follows from Equation (78) in Mehra \textit{et al.} \parencite{mehra2022hypnozoite} and $P(N_H(t)=0|N_A(t)=N_P(t)=0)$ is obtained by dividing Equation (\ref{eqn:nHAP0}) by Equation (\ref{eqn:nAP0}).

\subsection{Average number of hypnozoites within liver-stage infected individuals:~$k_T(t)$}\label{sec:k_T}

The average number of hypnozoites within liver-stage infected individuals, $k_T(t)$, is defined in Anwar \textit{et al.} 2023 \cite{anwar2023optimal} as
\begin{align}
\label{eqn:inf_sum}
   k_T=\sum_{i=1}^\infty ik_i &= \Big(\frac{\mathbb{E}\left[N_H(t)|N_A(t)=N_P(t)=0\right]}{1-P(N_H(t)=0|N_A(t)=N_P(t)=0)}\Big),\nonumber
\end{align}
where $\mathbb{E}\left[N_H(t)|N_A(t)=N_P(t)=0\right]$ is the expected size of the hypnozoite reservoir in an uninfected (no blood-stage infection) individual under $N$ rounds of MDA and is given by
\begin{align}
   &\mathbb{E}\left[N_H(t)|N_A(t)=N_P(t)=0\right]
  \nonumber \\
   =&\begin{cases} 
   \int_0^t \frac{\nu p_H(t-\tau)\lambda(\tau)\big(1-e^{-\gamma (t-\tau)}\big) }{[1+\nu p_A(t-\tau)]^2 }d\tau & \text{if}\ t< s_1,
   \\
   \int_{s_1}^t \frac{\nu p_H(t-\tau)\lambda(\tau)\big(1-e^{-\gamma (t-\tau)}\big) }{[1+\nu p_A(t-\tau)]^2 }d\tau \\ \quad+
   \int_{0}^{s_1} \frac{\nu p_H^r(t-\tau,s_1-\tau)\lambda(\tau)\big(1-(1-p_{\text{blood}})e^{-\gamma (t-\tau)}\big) }{[1+\nu p_A^r(t-\tau,s_1-\tau)]^2 }d\tau\\\quad+
   \int_{s_{1}}^{s_2} \frac{\nu p_H^r(t-\tau,s_1-\tau,s_2-\tau)\lambda(\tau)\big(1-(1-p_{\text{blood}})^2e^{-\gamma (t-\tau)}\big) }{[1+\nu p_A^r(t-\tau,s_1-\tau,s_2-\tau)]^2 }d\tau & \text{if}\ s_1 \le t< s_2,
  \\ 
    \quad \vdots 
    \\
   \int_{s_{N_{\text{MDA}}}}^t \frac{\nu p_H(t-\tau)\lambda(\tau)\big(1-e^{-\gamma (t-\tau)}\big) }{[1+\nu p_A(t-\tau)]^2 }d\tau \\ \quad+
   \int_{0}^{s_1} \frac{\nu p_H^r(t-\tau,s_1-\tau)\lambda(\tau)\big(1-(1-p_{\text{blood}})e^{-\gamma (t-\tau)}\big) }{[1+\nu p_A^r(t-\tau,s_1-\tau)]^2 }d\tau\\\quad+
   \int_{s_{1}}^{s_2} \frac{\nu p_H^r(t-\tau,s_1-\tau,s_2-\tau)\lambda(\tau)\big(1-(1-p_{\text{blood}})^2e^{-\gamma (t-\tau)}\big) }{[1+\nu p_A^r(t-\tau,s_1-\tau,s_2-\tau)]^2 }d\tau \\ \quad+\ldots +
   \int_{s_{N-1}}^{s_{N_{\text{MDA}}}} \frac{\nu p_H^r(t-\tau,s_1-\tau,\ldots,s_{N_{\text{MDA}}}-\tau)\lambda(\tau)\big(1-(1-p_{\text{blood}})^{N_{\text{MDA}}}e^{-\gamma (t-\tau)}\big) }{[1+\nu p_A^r(t-\tau,s_1-\tau,\ldots,s_{N_{\text{MDA}}}-\tau)]^2 }d\tau & \text{if}\ t\ge s_{N_{\text{MDA}}}.
\end{cases}
\end{align}

 \section{Recovery with superinfection}

When considering superinfection, an individual experiences multiple (at least one) blood-stage infections (either from a new infectious bite or hypnozoite activation) at the same time. We define the multiplicity of infection ($MOI$) as the number of distinct parasite broods co-circulating within a blood-stage infected individual from different infections. Thus, $MOI$ is given by the total number of bloods-stage infections (infections from mosquito bites and relapses) at time $t$: $M_I(t)=N_A(t)+N_P(t)$, where $N_A$ and $N_P$ represent the number of relapse and primary infections at time $t$. Thus, the probability of recovery from a blood-stage infection ($I$) is conditional upon how many infections ($MOI$) they are currently experiencing. As per Mehra \etal\parencite{thesis_somya} and Nåsell \textit{et al.} \parencite{nasell2013hybrid}, the recovery rate under superinfection is 
\begin{align}
\label{eqn:reco}
    \text{Recovery rate}=&\gamma P(N_A(t)+N_P(t)=1|P(N_A(t)+N_P(t)>0),\nonumber
     \\=&\frac{\gamma P(N_A(t)+N_P(t)=1)}{P(N_A(t)+N_P(t)>0)},
\end{align}
where $\gamma$ is the recovery rate without superinfection. The recovery rate is conditioned upon having at least one infection at a time $t$, and after the recovery, individuals move out of the blood-stage infected compartment ($I$). Previously, in our model without superinfection (Anwar \textit{et al.} 2022 \cite{anwar2022multiscale}), individuals from the blood-stage infected compartment ($I$) compartment move to the susceptible compartment ($S$) at rate $\gamma p(t)$ and move to the liver-stage infected compartment ($L$) at rate $\gamma (1-p(t))$ because of the assumption of a constant recovery rate, $\gamma$. That is, we only needed the hypnozoite distribution given an individual in the blood-stage infected compartment. However, with superinfection, individuals from the blood-stage infected compartment ($I$) will move back to the susceptible compartment ($S$) if they are experiencing only one infection and have no hypnozoites. That is, the recovery rate depends upon the multiplicity of infection and hypnozoite status, which means the probability $p(t)$ is insufficient to characterise the dynamics. Therefore, we need the joint probability of experiencing only one infection together with the hypnozoite reservoir size. 

\subsection{Probability of blood-stage infected individual having one infection and no hypnozoites:~$p_1(t)$}

Following the work of Mehra \cite{thesis_somya}, we define $p_1(t)$ to be the probability that a blood-stage infected individual has no hypnozoites and is experiencing only one infection at time $t$. That is,
\begin{align}p_1(t)=&P(N_A(t)+N_P(t)=1,N_H(t)=0|P(N_A(t)+N_P(t)>0),\nonumber\\
             =&\frac{P\big(N_A(t)+N_P(t)=1\ ,N_H(t)=0\big)}{P(N_A(t)+N_P(t)>0)},\nonumber\\
             =&\frac{P\big(N_A(t)+N_P(t)=1|N_H(t)=0\big)P(N_H(t)=0)}{1-P(N_A(t)+N_P(t)=0)},
             \label{eqn:ch5/p_1}
\end{align}

where the expression for $P(N_H(t)=0)$ and $P(N_A(t)+N_P(t)=0)$
follows from Equations (\ref{eqn:nH0}) and (\ref{eqn:nAP0}).

Now, multiplicity of infection given an empty hypnozoite reservoir, $P(N_A(t)+N_P(t)|N_H(t)=0)$, can be obtained from the PGF given by Equation (\ref{PGF_ntrt}) that holds for before treatment and by Equation (\ref{PGF}) which holds following treatment at times $t=s_1,\ s_2,\ \ldots, s_{N_{\text{MDA}}}$ as

\begin{align*}
  \mathbb{E}[z^{M_I(t)}|N_H(t)=0]=&\begin{cases}
    \frac{G(t,z_H=0,\ z_A=z,\ z_C=1,\ z_D=1,\ z_P=z\, z_{PC}=1)}{G(t,z_H=0,\ z_A=1,\ z_C=1,\ z_D=1,\ z_P=1\, z_{PC}=1)} & \text{if}\ t< s_1
    \\
    \frac{G^{s_1}(t,z_H=0,\ z_A=z,\ z_C=1,\ z_D=1,\ z_P=z\, z_{PC}=1)}{G^{s_1}(t,z_H=0,\ z_A=1,\ z_C=1,\ z_D=1,\ z_P=1\, z_{PC}=1)} & \text{if}\ s_1 \le t< s_2,
  \\ 
    \quad \vdots 
    \\
    \frac{G^{s_1,s_2,\ldots s_{N_{\text{MDA}}}}(t,z_H=0,\ z_A=z,\ z_C=1,\ z_D=1,\ z_P=z\, z_{PC}=1)}{G^{s_1,s_2,\ldots s_{N_{\text{MDA}}}}(t,z_H=0,\ z_A=1,\ z_C=1,\ z_D=1,\ z_P=1\, z_{PC}=1)} & \text{if}\ t\ge s_{N_{\text{MDA}}},\\
    \end{cases}\\
    =&\text{exp}\{h(z,t)-h(1,t)\},
\end{align*}

where
\begin{align}
   h(z,t)=&\begin{cases} 
   \int_0^t \lambda(\tau)\frac{ze^{-\gamma (t-\tau)}+(1-e^{-\gamma (t-\tau)})}{1+\big(p_H(t-\tau)+(1-z)p_A(t-\tau)\big)\nu }d\tau & \text{if}\ t< s_1\\
   \int_{s_1}^t \lambda(\tau)\frac{ze^{-\gamma (t-\tau)}+(1-e^{-\gamma (t-\tau)})}{1+\big(p_H(t-\tau)+(1-z)p_A(t-\tau)\big)\nu }d\tau\\\quad+\int_0^{s_1} \lambda(\tau)\frac{z(1-p_{blood})e^{-\gamma (t-\tau)}+(1-(1-p_{blood})e^{-\gamma (t-\tau)})}{1+\big(p_H(t-\tau,s_1-\tau)+(1-z)p_A(t-\tau,s_1-\tau)\big)\nu }d\tau & \text{if}\ s_1 \le t< s_2,
  \\ 
    \quad \vdots 
    \\
    \int_{s_{N_{\text{MDA}}}}^t \lambda(\tau)\frac{ze^{-\gamma (t-\tau)}+(1-e^{-\gamma (t-\tau)})}{1+\big(p_H(t-\tau)+(1-z)p_A(t-\tau)\big)\nu }d\tau\\\quad+\int_0^{s_1} \lambda(\tau)\frac{z(1-p_{blood})e^{-\gamma (t-\tau)}+(1-(1-p_{blood})e^{-\gamma (t-\tau)})}{1+\big(p_H(t-\tau,s_1-\tau)+(1-z)p_A(t-\tau,s_1-\tau)\big)\nu }d\tau\\\quad+\ldots+\int_{s_{N-1}}^{s_{N_{\text{MDA}}}} \lambda(\tau)\frac{z(1-p_{blood})^{N_{\text{MDA}}}e^{-\gamma (t-\tau)}+(1-(1-p_{blood})^{N_{\text{MDA}}}e^{-\gamma (t-\tau)})}{1+\big(p_H(t-\tau,s_1-\tau,\ldots,s_{N_{\text{MDA}}}-\tau)+(1-z)p_A(t-\tau,s_1-\tau,\ldots,s_{N_{\text{MDA}}}-\tau)\big)\nu }d\tau
    & \text{if}\ t\ge s_{N_{\text{MDA}}}.
    \end{cases}\label{eqn:h}
\end{align}

Now, the PMF for $M_I(t)|N_H(t)=0$ is
\begin{align*}
% \label{eqn:81_new}
&P(N_A(t)+N_P(t)=n|N_H(t)=0)=P(M_I(t)=n|N_H(t)=0),\nonumber\\
=&\text{exp}\left\{h(0,t)-h(1,t)\right\}\frac{1}{n!}\sum_{k=1}^n B_{n,k}\left[\frac{\partial h(0,t)}{\partial z},\frac{\partial^2 h(0,t)}{\partial z^2},\ldots,\frac{\partial^{n-k+1} h(0,t)}{\partial z^{n-k+1}}\right],
\end{align*}
where
\begin{align*}
   \frac{\partial^{k} h}{\partial z^k}(0,t)=&\begin{cases}
    k!\int_0^t \frac{\lambda(\tau)[\nu p_A(t-\tau)]^{k-1}}{\big[1+\nu \big(p_A(t-\tau)+p_H(t-\tau)\big)\big]^k}\Big(e^{-\gamma (t-\tau)}+\frac{\nu p_A(t-\tau)(1-e^{-\gamma (t-\tau)})}{1+\nu p_A(t-\tau)}\Big)d\tau& \text{if}\ t< s_1
    \\
    k!\Big(\int_{s_1}^t \frac{\lambda(\tau)\nu p_A(t-\tau)^{k-1}}{\big[1+\nu \big(p_A(t-\tau)+p_H(t-\tau)\big)\big]^k}\Big(e^{-\gamma (t-\tau)}+\frac{\nu p_A(t-\tau)(1-e^{-\gamma (t-\tau)})}{1+\nu p_A(t-\tau)}\Big)d\tau \\\quad+\int_0^{s_1} \frac{\lambda(\tau)\nu {p_A^r(t-\tau,s_1-\tau)}^{k-1}}{\big[1+\nu \big(p_A^r(t-\tau,s_1-\tau)+p_H^r(t-\tau,s_1-\tau)\big)\big]^k}\Big((1-p_{blood})e^{-\gamma (t-\tau)}\\\quad+\frac{\nu{p_A^r(t-\tau,s_1-\tau)}(1-(1-p_{blood})e^{-\gamma (t-\tau)})}{1+\nu {p_A^r(t-\tau,s_1-\tau)}}\Big)d\tau & \text{if}\ s_1 \le t< s_2,
  \\ 
    \quad \vdots 
    \\
    k!\Big(\int_{s_{N_{\text{MDA}}}}^t \frac{\lambda(\tau)\nu p_A(t-\tau)^{k-1}}{\big[1+\nu \big(p_A(t-\tau)+p_H(t-\tau)\big)\big]^k}\Big(e^{-\gamma (t-\tau)}+\frac{\nu p_A(t-\tau)(1-e^{-\gamma (t-\tau)})}{1+\nu p_A(t-\tau)}\Big)d\tau \\\quad+\int_0^{s_1} \frac{\lambda(\tau)\nu {p_A^r(t-\tau,s_1-\tau)}^{k-1}}{\big[1+\nu \big(p_A^r(t-\tau,s_1-\tau)+p_H^r(t-\tau,s_1-\tau)\big)\big]^k}\Big((1-p_{blood})e^{-\gamma (t-\tau)}\\\quad+\frac{\nu{p_A^r(t-\tau,s_1-\tau)}(1-(1-p_{blood})e^{-\gamma (t-\tau)})}{1+\nu {p_A^r(t-\tau,s_1-\tau)}}\Big)d\tau+\ldots\\\quad+\int_{s_{N-1}}^{s_{N_{\text{MDA}}}} \frac{\lambda(\tau)\nu {p_A^r(t-\tau,s_1-\tau,\ldots,s_{N_{\text{MDA}}}-\tau)}^{k-1}}{\big[1+\nu \big(p_A^r(t-\tau,s_1-\tau,\ldots,s_{N_{\text{MDA}}}-\tau)+p_H^r(t-\tau,s_1-\tau,\ldots,s_{N_{\text{MDA}}}-\tau)\big]^k}\\\quad\Big((1-p_{blood})^Ne^{-\gamma (t-\tau)}+\frac{\nu {p_A^r(t-\tau,s_1-\tau,\ldots,s_{N_{\text{MDA}}}-\tau)}(1-(1-p_{blood})^Ne^{-\gamma (t-\tau)})}{1+\nu {p_A^r(t-\tau,s_1-\tau,\ldots,s_{N_{\text{MDA}}}-\tau)}}\Big)d\tau\Big)
    & \text{if}\ t\ge s_{N_{\text{MDA}}}.\\
    \end{cases}
\end{align*}

Therefore, 
\begin{align*}
% \label{eqn:k1}
P(N_A(t)+N_P(t)=&1|N_H(t)=0)=\text{exp}\left\{h(0,t)-h(1,t)\right\}\frac{\partial h(0,t)}{\partial z},\\
=&\frac{G(t,z_H=0,\ z_A=0,\ z_C=1,\ z_D=1,\ z_P=0\, z_{PC}=1)}{G(t,z_H=0,\ z_A=1,\ z_C=1,\ z_D=1,\ z_P=1\, z_{PC}=1)}\frac{\partial h(0,t)}{\partial z},\\
=&\frac{P(N_H(t)=N_A(t)=N_P(t)=0)}{P(N_H(t)=0)}\frac{\partial h(0,t)}{\partial z}.
 \end{align*}
Finally, from Equation (\ref{eqn:ch5/p_1})
\begin{align}
\label{eqn:p_1_final}       p_1(t)=&\frac{P\big(N_A(t)+N_P(t)=1|N_H(t)=0\big)P(N_H(t)=0)}{1-P(N_A(t)+N_P(t)=0)},\nonumber\\
        =&\frac{P(N_H(t)=N_A(t)=N_P(t)=0)}{1-P(N_A(t)=N_P(t)=0)}\frac{\partial h(0,t)}{\partial z},
\end{align}
where 
\begin{align*}
% \label{eqn:k1}
\frac{\partial h(0,t)}{\partial z}=&\begin{cases}
\int_0^t \lambda(\tau)\frac{e^{-\gamma (t-\tau)}\big(1+\nu p_H(t-\tau)\big)+\nu p_A(t-\tau)}{\big[1+\nu\big(p_A(t-\tau)+p_H(t-\tau)\big)\big]^2}d\tau & \text{if}\ t< s_1,
\\
\int_{s_1}^t \lambda(\tau)\frac{e^{-\gamma (t-\tau)}\big(1+\nu p_H(t-\tau)\big)+\nu p_A(t-\tau)}{\big[1+\nu\big(p_A(t-\tau)+p_H(t-\tau)\big)\big]^2}d\tau \\\quad+\int_0^{s_1} \lambda(\tau)\frac{(1-p_{\text{blood}})e^{-\gamma (t-\tau)}\big(1+\nu p_H^r(t-\tau,s_1-\tau)\big)+\nu p_A^r(t-\tau,s_1-\tau)}{\big[1+\nu\big(p_A^r(t-\tau,s_1-\tau)+p_H^r(t-\tau,s_1-\tau)\big)\big]^2}d\tau \\\quad +\int_{s_{1}}^{s_2} \lambda(\tau)\frac{(1-p_{\text{blood}})^2e^{-\gamma (t-\tau)}\big(1+\nu p_H^r(t-\tau,s_1-\tau,s_2-\tau)\big)+\nu p_A^r(t-\tau,s_1-\tau,s_2-\tau)}{\big[1+\nu\big(p_A^r(t-\tau,s_1-\tau,s_{N_{\text{MDA}}}-\tau)+p_H^r(t-\tau,s_1-\tau,s_{N_{\text{MDA}}}-\tau)\big)\big]^2}d\tau\\ & \text{if}\ s_1 \le t< s_2,
  \\ 
    \quad \vdots 
    \\
    \int_{s_{N_{\text{MDA}}}}^t \lambda(\tau)\frac{e^{-\gamma (t-\tau)}\big(1+\nu p_H(t-\tau)\big)+\nu p_A(t-\tau)}{\big[1+\nu\big(p_A(t-\tau)+p_H(t-\tau)\big)\big]^2}d\tau \\\quad+\int_0^{s_1} \lambda(\tau)\frac{(1-p_{\text{blood}})e^{-\gamma (t-\tau)}\big(1+\nu p_H^r(t-\tau,s_1-\tau)\big)+\nu p_A^r(t-\tau,s_1-\tau)}{\big[1+\nu\big(p_A^r(t-\tau,s_1-\tau)+p_H^r(t-\tau,s_1-\tau)\big)\big]^2}d\tau \\\quad +\int_{s_{1}}^{s_2} \lambda(\tau)\frac{(1-p_{\text{blood}})^2e^{-\gamma (t-\tau)}\big(1+\nu p_H^r(t-\tau,s_1-\tau,s_2-\tau)\big)+\nu p_A^r(t-\tau,s_1-\tau,s_2-\tau)}{\big[1+\nu\big(p_A^r(t-\tau,s_1-\tau,s_{N_{\text{MDA}}}-\tau)+p_H^r(t-\tau,s_1-\tau,s_{N_{\text{MDA}}}-\tau)\big)\big]^2}d\tau\\\quad+\ldots
    +\int_{s_{N-1}}^{s_{N_{\text{MDA}}}} \lambda(\tau)\Big[\frac{(1-p_{\text{blood}})^{N_{\text{MDA}}}e^{-\gamma (t-\tau)}\big(1+\nu p_H^r(t-\tau,s_1-\tau,\ldots,s_{N_{\text{MDA}}}-\tau)\big)}{\big[1+\nu\big(p_A^r(t-\tau,s_1-\tau,\ldots,s_{N_{\text{MDA}}}-\tau)+p_H^r(t-\tau,s_1-\tau,\ldots,s_{N_{\text{MDA}}}-\tau)\big)\big]^2}\\\quad+\frac{\nu p_A^r(t-\tau,s_1-\tau,\ldots,s_{N_{\text{MDA}}}-\tau)}{\big[1+\nu\big(p_A^r(t-\tau,s_1-\tau,\ldots,s_{N_{\text{MDA}}}-\tau)+p_H^r(t-\tau,s_1-\tau,\ldots,s_{N_{\text{MDA}}}-\tau)\big)\big]^2}\Big]d\tau & \text{if}\ t\ge s_{N_{\text{MDA}}}.
    \end{cases}
\end{align*}

\subsection{Probability of blood-stage infected individual having one infection and  non-zero hypnozoites:~$p_2(t)$}

 Meanwhile, individuals in $I$ move to $L$ after recovery if they have hypnozoites. We define $p_2(t)$ to be the probability that a blood-stage infected individual has hypnozoites and is experiencing more than one blood-stage infection at time $t$. That is, 
\begin{align}
    p_2(t)=&P(N_A(t)+N_P(t)=1,N_H(t)>0|P(N_A(t)+N_P(t)>0),\nonumber\\
             =&\frac{P\big(N_A(t)+N_P(t)=1\ ,N_H(t)>0\big)}{P(N_A(t)+N_P(t)>0)},\nonumber\\
             =&\frac{P\big(N_A(t)+N_P(t)=1|N_H(t)>0\big)P(N_H(t)>0)}{1-P(N_A(t)+N_P(t)=0)},\nonumber\\
             =&\frac{P\big(N_A(t)+N_P(t)=1\big)-P\big(N_A(t)+N_P(t)=1|N_H(t)=0\big)P(N_H(t)=0)}{1-P(N_A(t)+N_P(t)=0)},\nonumber\\
             =&\frac{P\big(N_A(t)+N_P(t)=1\big)}{1-P(N_A(t)=N_P(t)=0)}-p_1(t).\label{eqn:ch5/p_2}
\end{align}

The expression $P(N_A(t)+N_P(t)=0)=P(N_A(t)=N_P(t)=0)$ is given by Equation (\ref{eqn:nAP0}). The expression for $P(N_A(t)+N_P(t)=1)$ follows from Equation (81) in Mehra \textit{et al.} \cite{mehra2022hypnozoite} and is given by
\begin{align*}
% \label{eqn:k1}
P(N_A(t)+N_P(t)=1)=&P(N_A(t)=N_P(t)=0)\frac{\partial f(0,t)}{\partial z},\\
% =&\text{exp}\left\{-q(t)+\int_0^t  \frac{\lambda(\tau)(1-e^{-\gamma (t-\tau)})}{1+\nu P_A(t-\tau)}\right\}\frac{\partial f_1(0,t)}{\partial z},\\
%  =&P(N_A(t)=N_P(t)=0)
    % \frac{\partial f(0,t)}{\partial z},
 \end{align*}
where
\begin{align*}
% \label{eqn:k1}
\frac{\partial f(0,t)}{\partial z}=&\begin{cases}
\int_0^t \lambda(\tau)\frac{e^{-\gamma (t-\tau)}+\nu p_A(t-\tau)}{[1+\nu p_A(t-\tau)]^2}d\tau & \text{if}\ t< s_1,
\\
\int_{s_1}^t \lambda(\tau)\frac{e^{-\gamma (t-\tau)}+\nu p_A(t-\tau)}{[1+\nu p_A(t-\tau)]^2}d\tau \\\quad+\int_0^{s_1} \lambda(\tau)\frac{(1-p_{\text{blood}})e^{-\gamma (t-\tau)}+\nu p_A^r(t-\tau,s_1-\tau)}{[1+\nu p_A^r(t-\tau,s_1-\tau)]^2}d\tau\\\quad+\int_{s_{1}}^{s_2} \lambda(\tau)\frac{(1-p_{\text{blood}})^2e^{-\gamma (t-\tau)}+\nu p_A^r(t-\tau,s_1-\tau,s_2-\tau)}{[1+\nu p_A^r(t-\tau,s_1-\tau,s_2-\tau)]^2}d\tau \\\quad +\ldots+\int_{s_{N-1}}^{s_{N_{\text{MDA}}}} \lambda(\tau)\frac{(1-p_{\text{blood}})^{N_{\text{MDA}}}e^{-\gamma (t-\tau)}+\nu p_A^r(t-\tau,s_1-\tau,\ldots,s_{N_{\text{MDA}}}-\tau)}{[1+\nu p_A^r(t-\tau,s_1-\tau,\ldots,s_{N_{\text{MDA}}}-\tau)]^2}d\tau & \text{if}\ s_1 \le t< s_2,
  \\ 
    \quad \vdots\\
    \int_{s_{N_{\text{MDA}}}}^t \lambda(\tau)\frac{e^{-\gamma (t-\tau)}+\nu p_A(t-\tau)}{[1+\nu p_A(t-\tau)]^2}d\tau \\\quad+\int_0^{s_1} \lambda(\tau)\frac{(1-p_{\text{blood}})e^{-\gamma (t-\tau)}+\nu p_A^r(t-\tau,s_1-\tau)}{[1+\nu p_A^r(t-\tau,s_1-\tau)]^2}d\tau\\\quad+\int_{s_{1}}^{s_2} \lambda(\tau)\frac{(1-p_{\text{blood}})^2e^{-\gamma (t-\tau)}+\nu p_A^r(t-\tau,s_1-\tau,s_2-\tau)}{[1+\nu p_A^r(t-\tau,s_1-\tau,s_2-\tau)]^2}d\tau \\\quad +\ldots
    +\int_{s_{N-1}}^{s_{N_{\text{MDA}}}} \lambda(\tau)\frac{(1-p_{\text{blood}})^{N_{\text{MDA}}}e^{-\gamma (t-\tau)}+\nu p_A^r(t-\tau,s_1-\tau,\ldots,s_{N_{\text{MDA}}}-\tau)}{[1+\nu p_A^r(t-\tau,s_1-\tau,\ldots,s_{N_{\text{MDA}}}-\tau)]^2}d\tau & \text{if}\ t\ge s_{N_{\text{MDA}}}.
    \end{cases}
\end{align*}

The time-dependent probabilities $p(t),\ p_1(t),\ p_2(t),\ k_1(t)$, and $k_T(t)$ that characterise the hypnozoite dynamics at the population level, account for all the infective bites received throughout time (as a function of $\lambda(t)$) and change instantaneously with MDA because of the assumption of the instantaneous effect of the drug.

% {\color{purple}AEZ: The references are missing DOIs. Possibly not essential, but it definitely makes it easier to follow up on papers if they are there. If not wildly difficult, please include hyperlinked DOIs.}

\begin{landscape}
\begin{table}[ht]
\centering
\caption{Optimal interval for up to four rounds of MDA obtained from the deterministic version of the model.}
\begin{center}
\begin{tabular}{|c|c|c|c|c|c|c|c|c|c|c|c|}
    \hline
\multicolumn{3}{|c|}{Prevalence} &10$\%$& 20$\%$&30$\%$&40$\%$&50$\%$ & 60$\%$& 70$\%$&80$\%$& 90$\%$\\
\hline
  \multirow{10}{*}{Interval (days)} &
  \multirow{1}{*}{One round} &$x_0$ (Peak to MDA1)&116.4&116.4&116.4&102.4&102.4&102.4&69&69&58.7\\
  \cline{2-12}
& \multirow{2}{*}{Two rounds} &$x_0$ (Peak to MDA1) &139.9&139.9&139.9&139.9&139.9&139.9&101.8&101.8&101.8\\
\cline{3-12}
& &$x_1$ (MDA1 to MDA2) &28.6&28.6&28.6&28.6&28.6&28.6&303.4&303.4&303.4 \\
% \cline{3-11}
% & & Set &&&&&&&& \\
\cline{2-12}
& \multirow{3}{*}{Three rounds} & $x_0$ (Peak to MDA1) &10&139.9&139.9&139.9&139.9&139.9&116.4&116.4&116.4\\
\cline{3-12}
& &$x_1$ (MDA1 to MDA2) &10&28.6&28.6&28.6&28.6&28.6&116.4&116.4&116.4 \\
\cline{3-12}
& &$x_2$ (MDA2 to MDA3) &10&53.8&53.8&53.8&53.8&53.8&50.3&50.3&50.3 \\
\cline{2-12}
& \multirow{4}{*}{Four rounds} & $x_0$ (Peak to MDA1) &139.9&139.9&139.9&139.9&139.9&139.9&116.4&116.4&52.7\\
\cline{3-12}
& &$x_1$ (MDA1 to MDA2) &28.6&28.6&28.6&28.6&28.6&28.6&276.6&276.6&325.3 \\
\cline{3-12}
& &$x_2$ (MDA2 to MDA3) &53.8&53.8&53.8&53.8&53.8&53.8&50.3&50.3&333.5 \\ 
\cline{3-12}
& &$x_3$ (MDA3 to MDA4) &285&285&285&285&285&285&285&285&88.9 \\ 
\hline
\end{tabular}
\end{center}
\label{tab:optimal_interval}
\end{table}
\end{landscape}

\begin{figure}[!ht]
\centering
  \includegraphics[width=\textwidth]{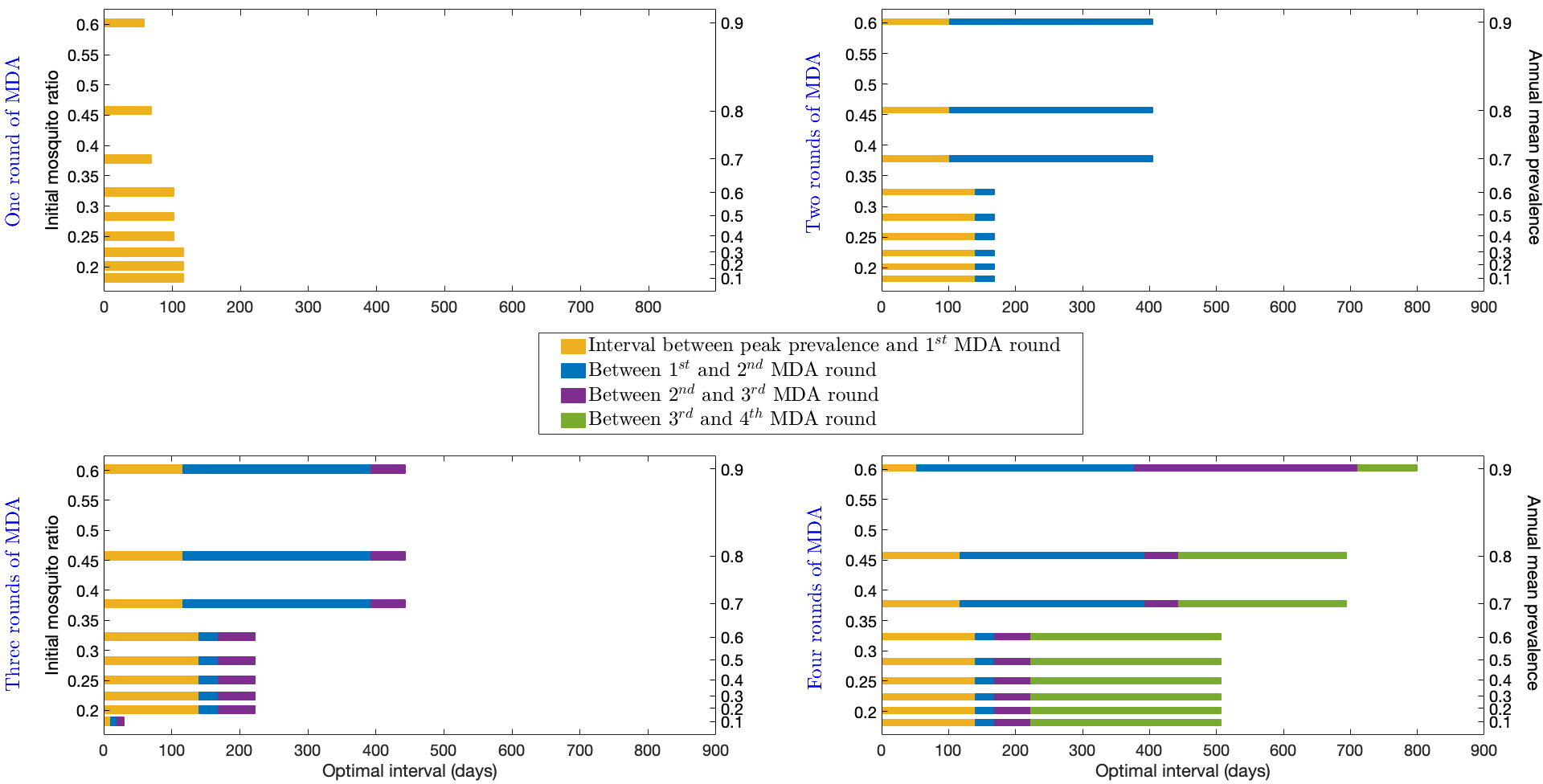}
   \caption{\textit{Optimal MDA intervals for up to four rounds obtained with the deterministic version of the model for varying mosquito to human $m_0$  (left vertical axis). These different initial mosquito ratios correspond to the prevalence of blood-stage infection in the range of 10--90\% (right vertical axis).}}
  \label{fig:sup_MDA1_MDA4}
\end{figure}

\end{document}